\documentclass[aps,prx,reprint,superscriptaddress,floatfix]{revtex4-2}
\usepackage{amsmath,amssymb,amsfonts}
\usepackage{slashed}
\usepackage{graphicx}
\usepackage{bm}
\usepackage{float} 
\usepackage{needspace}
\usepackage{hyperref}
\hypersetup{hidelinks}
\usepackage{booktabs}

\newcommand{\Kem}{K_{\rm env}}
\newcommand{\Dpg}{\Delta_{\rm pg}}
\newcommand{\Dsc}{\Delta_{\rm sc}}
\newcommand{\Tstar}{T^{*}}

\newcommand{\vth}{\vartheta}
\newcommand{\Leff}{L_{\rm eff}}
\newcommand{\R}{\mathbb{R}}
\newcommand{\dd}{\mathrm{d}}
\newcommand{\tr}{\mathrm{tr}}
\newcommand{\sgn}{\mathrm{sgn}}
\newcommand{\CBdG}{C_{\rm BdG}}

\begin{document}

\title{Thermal Hall tomography of superconducting topology\\ in rhombohedral graphene}

\author{Kumar Ghosh}
\email{jb.ghosh@outlook.com}
\affiliation{E.ON Digital Technology, Hannover, Germany}

\begin{abstract}
Chiral superconductors should carry Majorana modes along their edges,
yet three decades of work on candidate materials have not measured the
integer that counts them. Rhombohedral graphene offers a new route,
because its superconductivity coexists with magnetic domains that can
be imaged and rewritten. Here we show how heat transport can read this
integer and test it against the domain pattern. Warping and finite pair
momentum can close the excitation gap while the band Chern number stays
fixed. An exact two-band thermal kernel shows how quantization is then
lost and bounds the plateau error when the gap survives. Real-space
calculations with disorder show that the integer survives disorder
several times larger than the gap, because a mobility gap remains.
Bogoliubov pockets instead turn the bulk into a heat conductor, and
disorder does not restore quantization. Time-reversed domains should
show opposite plateaux, and with ideal contacts a wall between them
should carry twice either plateau. In microscopic calculations, a
written wall carries this number of Majorana channels even when its
core is metallic. Writing,
moving and erasing such walls would test superconducting topology, its
link to magnetism, and protected heat flow in one device.
\end{abstract}

\maketitle

\section{Introduction}
\label{sec:intro}

Chiral superconductors are a long-sought home for Majorana physics.
Their boundaries carry chiral Majorana modes, and the net number of
these modes is fixed by an integer, the superconducting Chern number.
For odd Chern number, an isolated vortex can also bind a Majorana zero
mode~\cite{Read2000,Kallin2016,SatoAndo2017}. Measuring this integer
has proved difficult. Magnetic signals show that time-reversal symmetry
is broken, but they do not count modes. Magnetization and equilibrium
edge currents depend on microscopic structure, disorder, and
screening, and they are not fixed by topology~\cite{Kallin2016}.

Heat avoids this ambiguity. Majorana modes carry no charge, but they
do carry energy. In a fully gapped superconductor the low-temperature
thermal Hall conductance is set by the integer alone. Quantized heat
flow has already resolved neutral edge modes in quantum Hall
systems~\cite{Banerjee2018}. What superconductors have lacked is a
control knob: a way to show that a measured plateau belongs to the
superconducting state and follows its chirality.

Rhombohedral graphene supplies such a knob. Its superconductivity
appears beside an orbital ferromagnet, with hysteretic transport and
large critical fields~\cite{Han2025}. Magnetic imaging resolves isospin
domains inside the superconducting phase, and small currents can
rewrite them~\cite{Dutta2026}. This raises a concrete question: can a
magnetic domain wall be written as a path for protected heat? Two
uncertainties stand in the way. The normal-state Fermi surface is still
debated, because quantum oscillations show several tones and possible
magnetic breakdown~\cite{Kalantre2026,Zhao2026}. Microscopic pairing
proposals also allow different topological
sectors~\cite{Geier2025,Patri2025,li2025berrytrashcan,MayMann2026}.

Here we develop \emph{thermal Hall tomography}. It combines three
measurements in one device: the topological integer, its reversal with
an imaged domain, and the heat channel along a written domain wall.
For a fully gapped two-dimensional class-D superconductor, the total
physical BdG Chern number $\CBdG$ fixes the low-temperature electronic
thermal Hall coefficient:
\begin{equation}
 \frac{\kappa_{xy}}{T}\bigg|_{T\to0}
 = \frac{\pi^2k_B^2}{6h}\,\CBdG,
\label{eq:kxy-intro-quantum}
\end{equation}
Here $\kappa_{xy}$ is a sheet thermal Hall conductance and
$q_M=\pi^2k_B^2/(6h)$ is one Majorana thermal quantum.
The relation follows from the chiral central charge
$c_-=\CBdG/2$~\cite{Read2000,Kane1997,Senthil1999,Ryu2012,SumiyoshiFujimoto2013}.
Its use requires a gapped bulk, controlled parasitic heat paths, and
contacts that equilibrate the relevant edge modes.

\begin{figure}[H]
  \setlength{\columnwidth}{\textwidth}
  \setlength{\hsize}{\textwidth}
  \setlength{\linewidth}{\textwidth}
  \centering
  \includegraphics[width=0.9\textwidth]{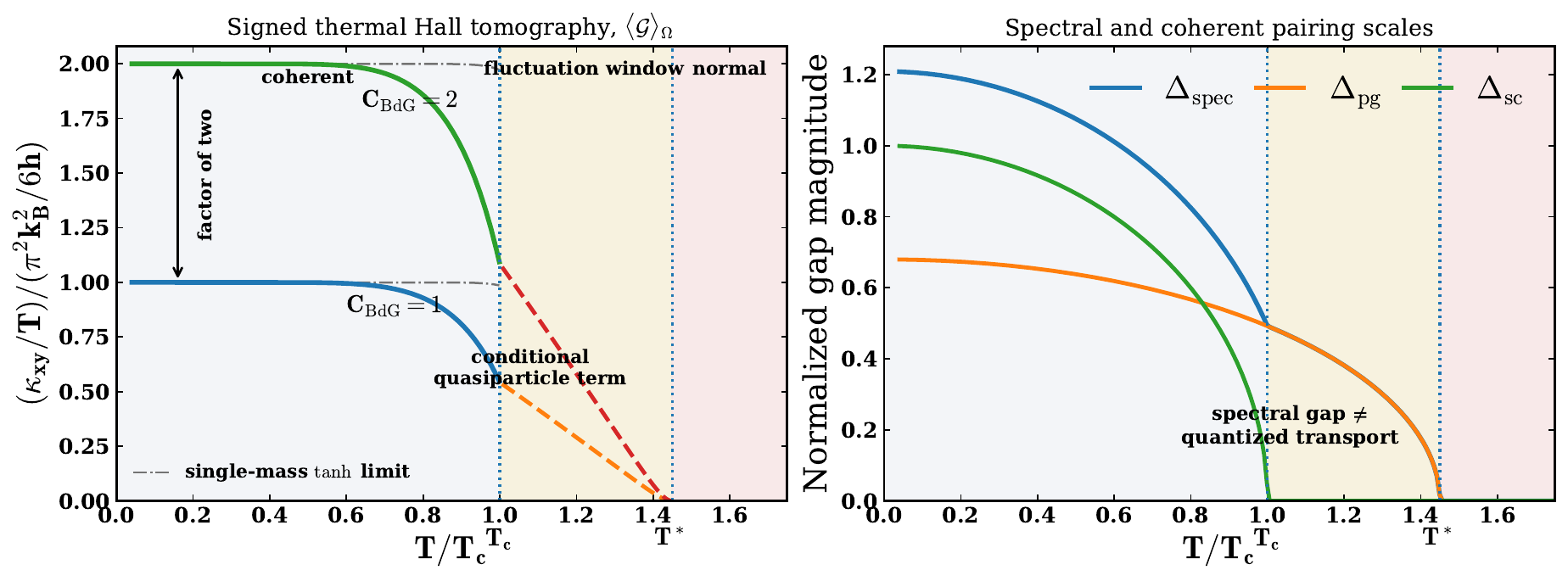}
  \caption{Illustrative finite-temperature framework for thermal Hall tomography.
(a) The signed curvature-weighted thermal kernel on a rescaled lattice
spectrum. The $C=2$ curve combines two identical $C=1$ contributions
for illustration. Only the low-temperature factor-of-two plateau is
fixed by the Chern number; the finite-temperature shape depends on the
spectrum. Solid curves describe the coherent model. Dashed curves
assume long-lived oriented quasiparticles above $T_c$ and do not give
the full response of that regime. Grey curves show the single-mass
comparison $\tanh[\Delta/(2k_BT)]$, with the same plateau normalization.
(b) Illustrative condensate and pseudogap profiles satisfying
$\Delta^2=\Dsc^2+\Dpg^2$. The scale $\Tstar=1.45T_c$ is chosen.
A spectral gap alone does not establish a static BdG invariant or
quantized transport above $T_c$.}
  \label{fig:fig1_schematic}
\end{figure}

Figure~\ref{fig:fig1_schematic} sets out the finite-temperature
picture. The low-temperature plateau records the net chirality. The
departure from the plateau shows how the excitation energies distribute
the Berry curvature, the geometric weight that drives the Hall
response. Solid curves illustrate this in a coherent BdG model. Dashed
curves ask a separate question about long-lived chiral correlations
above $T_c$. They are a conditional illustration, not a prediction of
quantized transport without phase coherence.

We establish when such a measurement reads topology, and what it reads
when it cannot. Five results support the protocol. First, warping and
finite pair momentum can close the zero-energy gap while the band Chern
number stays fixed, so topology and its thermal readout can separate.
Second, an exact thermal kernel turns this separation into numbers. It
gives the loss of quantization and a temperature window for a plateau
of chosen accuracy. Third, a single odd pairing component can produce
an even Chern number, so an even plateau does not by itself imply
several pairing components. Fourth, real-space calculations with
disorder show that the plateau survives as long as a mobility gap
remains, even when disorder states fill the spectral gap. Extended
Bogoliubov pockets instead make the bulk conduct heat, and disorder does
not restore the integer. Fifth, a written wall between time-reversed
domains carries $2|C|$ Majorana channels, even when its core is
metallic. Resistive walls of the kind seen in
experiment~\cite{Dutta2026} could therefore still guide quantized heat
along their length. Together these results give a quantitative
domain-wall test. Opposite gapped domains should show opposite
plateaux, and the wall between them should carry heat at twice either
plateau value. This linked response is the central experimental
prediction of this work.

\onecolumngrid
\twocolumngrid

\section{Results}
\label{sec:results}

We first establish when a band invariant can be read from heat
transport. We then test this condition in vortex models, quantify the
thermal response, and test both against disorder in real space. Finally
we formulate the domain-wall prediction and check it microscopically.

\subsection{When band topology produces a thermal plateau}
\label{subsec:gapped}

Intravalley pairing has momentum $2\bm K+\bm Q$. For $\bm k$ measured
from the valley centre $\bm K$, define~\cite{YangZhang2025}
\begin{align}
  \xi_s(\bm{k}) &= \tfrac{1}{2}\bigl[
    \varepsilon_{\bm{k}+\bm{Q}/2} + \varepsilon_{-\bm{k}+\bm{Q}/2}\bigr] - \mu,
    \nonumber\\
  \xi_a(\bm{k}) &= \tfrac{1}{2}\bigl[
    \varepsilon_{\bm{k}+\bm{Q}/2} - \varepsilon_{-\bm{k}+\bm{Q}/2}\bigr],
\label{eq:xi-sa}
\end{align}
These are the symmetric and antisymmetric parts of the dispersion.
The Hamiltonian is
\begin{equation}
  H_{\rm BdG}(\bm{k}) = \xi_a(\bm{k})\,\tau_0
  + \bigl[\xi_s\tau_z
    + \mathrm{Re}\,\Delta_{\bm{k}}\,\tau_x
    - \mathrm{Im}\,\Delta_{\bm{k}}\,\tau_y\bigr],
\label{eq:BdG-split}
\end{equation}
Its eigenvalues are $E_\pm=\xi_a\pm\eta$, with
$\eta=\sqrt{\xi_s^2+|\Delta_{\bm k}|^2}$. The identity term shifts both
energies equally. At fixed $\xi_s$ and $\Delta$ it leaves the
eigenvectors unchanged. The direct band separation is $2\eta$, so a
Chern number for the isolated lower band stays defined whenever
$\eta>0$. Warping and $\bm Q$ can still change $\xi_s$, the
self-consistent gap, and the parent-band wave functions. They are not
inert in general.

A quantized ground-state thermal response needs one more condition:
\begin{equation}
  \boxed{\;
  \delta_{\rm BdG} \equiv \min_{\bm{k}}\Bigl[
    \sqrt{\xi_s^2(\bm{k}) + |\Delta_{\bm{k}}|^2} - |\xi_a(\bm{k})|
  \Bigr] > 0. \;}
\label{eq:gapped-condition}
\end{equation}
When it holds, the lower band is filled at every momentum and its
Chern number is the ground-state invariant. Where $|\xi_a|>\eta$, both
bands are filled or both are empty. The edge of such a region is a
Bogoliubov Fermi contour. The band invariant can then survive while the
thermal response loses its quantization. We call this continuing
invariant $C_{\rm band}$. It equals $\CBdG$ only when the spectrum is
fully gapped. Figures keep the label $C_{\rm BdG}$ for the band
calculation.

Within a smooth band gauge and a specified ultraviolet completion, the
occupied-vortex rule reads~\cite{LeNir2026}
\begin{equation}
  \CBdG = \sum_{\bm{k}_i\,\in\,D_{\rm occ}} w_i,
\label{eq:CBdG-vortex-intro}
\end{equation}
For the two-band model, $D_{\rm occ}=D_s=\{\bm k:\xi_s(\bm k)<0\}$.
At finite $\bm Q$ this is the pairing-relevant set, not the
normal-state Fermi sea. The charges $w_i$ follow the orientation of
the numerical Chern convention given in Methods. If the zero-energy
gap is absent, the right side classifies $C_{\rm band}$. For a
physical projected Chern band, the parent Bloch basis and remote
occupied sectors must also be included. A bare $2\times2$ calculation
cannot replace that embedding.

\subsection{Vortices and the transport window}
\label{subsec:warped}

We first use a triangular-lattice dispersion with threefold warping
and an imposed odd pairing texture. The gap has a central zero and a
pair of zeros at finite momentum. The normal dispersion has $C_3$
symmetry, but the chosen vortex pair and a fixed direction of $\bm Q$
need not. Appendix~\ref{app:warped} defines the model. It isolates how a
nonreciprocal dispersion changes the transport window for a fixed
vortex configuration. It tests this mechanism and does not select the
pairing state of a particular graphene device.

\begin{figure}[H]
  \setlength{\columnwidth}{\textwidth}
  \setlength{\hsize}{\textwidth}
  \setlength{\linewidth}{\textwidth}
  \centering
  \includegraphics[width=0.84\textwidth]{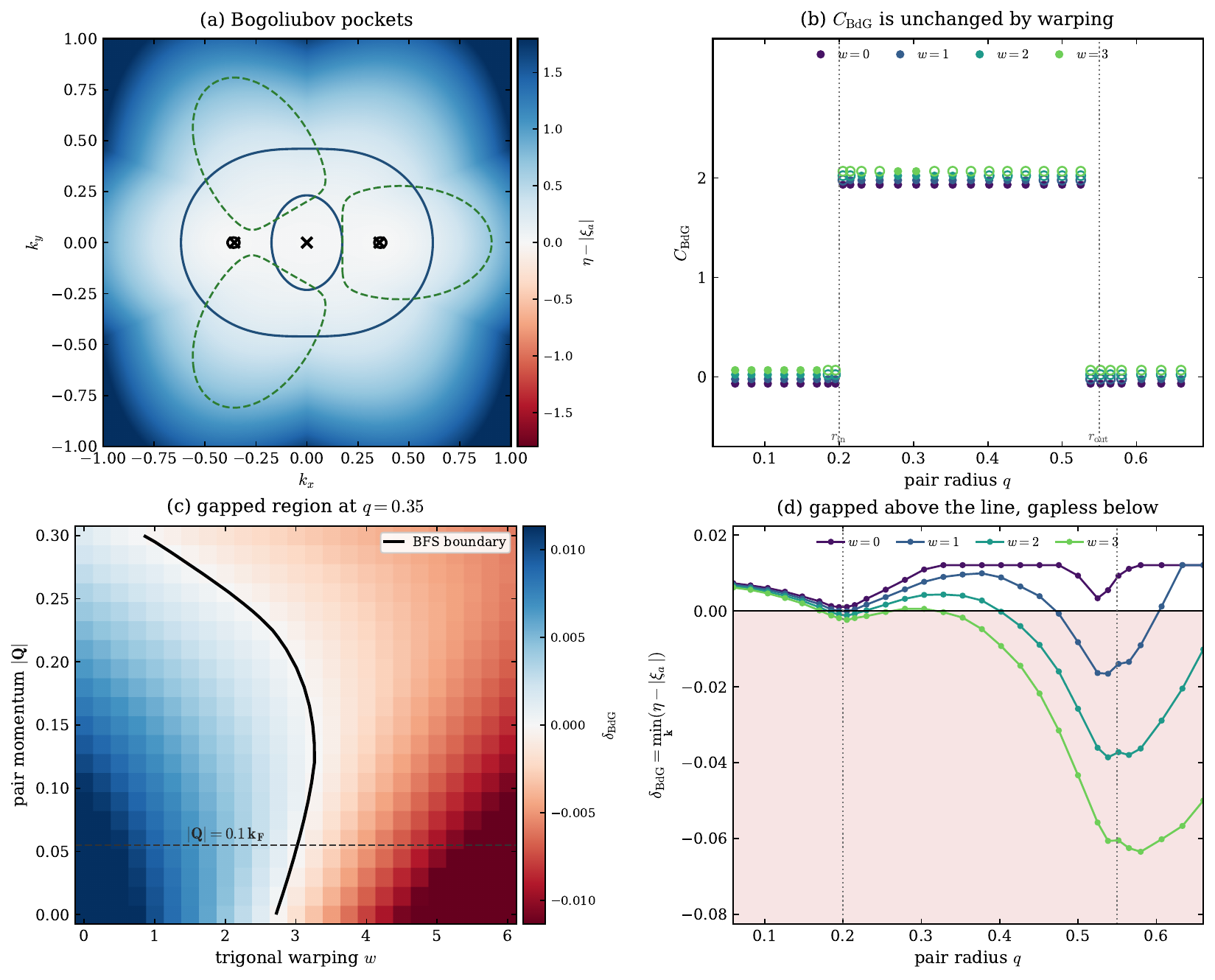}
  \caption{Band topology and the zero-energy gap in the imposed-vortex model.
(a) $\eta-|\xi_a|$ at $w=4$, $Q_x=0.10$, and $Q_y=0$.
Black contours mark zero energy. Blue contours show $\xi_s=0$, and
green dashed contours show the normal-state Fermi surface. Crosses
mark the primary gap zeros. Two gapless pockets form near the outer
zeros. (b) The vortex scan at four values of $w$, with vertical offsets.
Filled symbols are gapped and open symbols are gapless according to
the sampled minimum. (c) The gap estimate on the 525-point map at
$q=0.35$. The isolated-band Chern number is two at every point, while
the positive-gap fraction on this grid is $0.45$. Continuous refinement
can change the classification near the boundary. The horizontal
reference is $Q_x=0.1r_{\rm out}$; it is not gapped over the entire
warping range. (d) Gap estimates along the vortex scan. Direct-gap
crossings change the Chern number. Extended negative regions signal
Bogoliubov Fermi pockets without that change. In (b,d), dotted
$r_{\rm in,out}$ markers indicate continuum reference locations; the
actual lattice crossings are set by $\xi_s=0$ at the gap zeros. All
quantities are in model units. The full paired state need not have
$C_3$ symmetry.}
  \label{fig:warped}
\end{figure}

At $\bm Q=0$ the odd warping term cancels exactly from $\xi_s$, so the
Chern number does not depend on $w$. At finite $\bm Q$ the
cancellation is incomplete. Even so, a map with $w\in[0,6]$,
$Q_x\in[0,0.30]$, and $Q_y=0$ gives $C_{\rm band}=2$ at all 525
sampled points (Fig.~\ref{fig:warped}). On the finite grid,
$\delta_{\rm BdG}>0$ at only 45\% of these points. This fraction
depends on the mesh near the boundary. A stable band invariant thus
coexists with a large loss of the quantized-transport window.

\onecolumngrid
\twocolumngrid

The pockets form near the finite-momentum gap zeros inside the
occupied annulus. There the pairing is weak while the energy shift is
not. In the examples shown, two pockets appear and grow with warping.
Their centres move only slightly and are not pinned exactly to the
zeros. Near the boundary, continuous minimization is needed to resolve
small sampled gaps (Appendix~\ref{app:warped}). The reference momentum
$Q_x=0.1r_{\rm out}=0.055$ is motivated by Ref.~\cite{YangZhang2025};
its numerical threshold depends on the model.

\paragraph{An even invariant from one pairing component.}
\label{subsec:even-chern}

\begin{figure}[H]
  \setlength{\columnwidth}{\textwidth}
  \setlength{\hsize}{\textwidth}
  \setlength{\linewidth}{\textwidth}
  \centering
  \includegraphics[width=0.49\textwidth, height=0.35\textwidth]{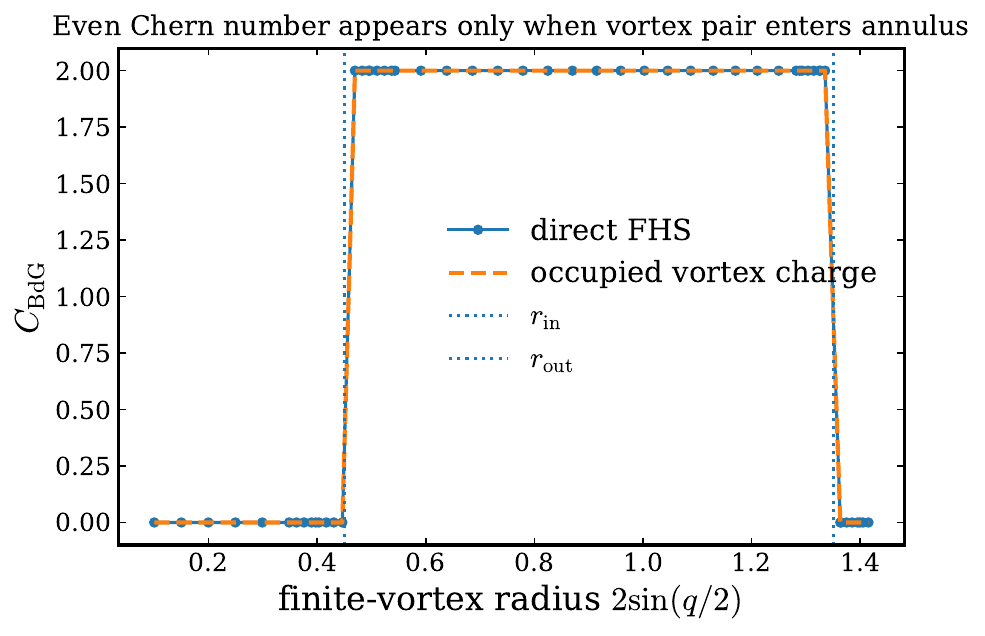}
  \hfill
  \includegraphics[width=0.49\textwidth, height=0.35\textwidth]{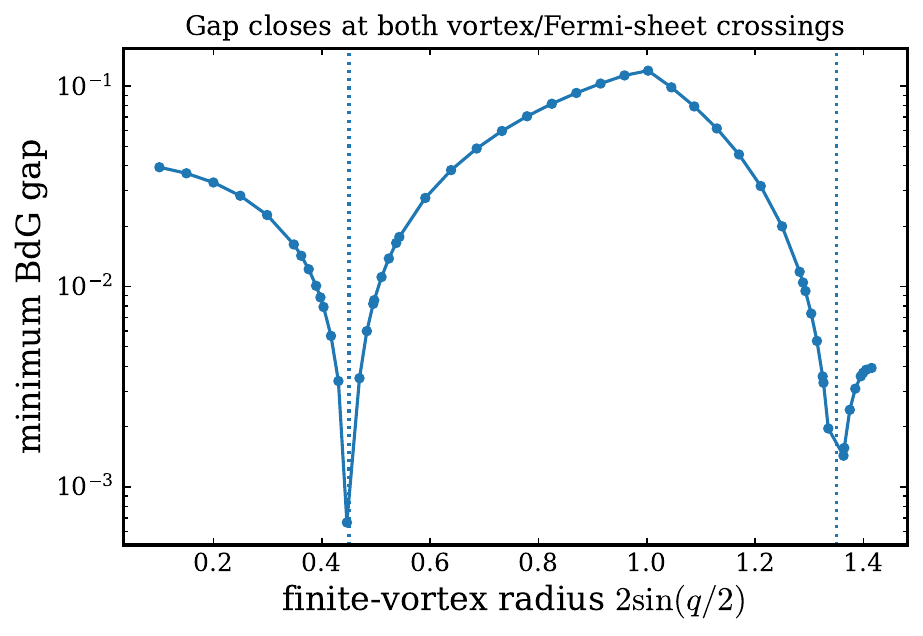}
  \caption{Annular vortex crossings and even Chern number in a single odd pairing component.
Left: the FHS invariant and the analytic oriented vortex charge over 54
primary pair radii $2\sin(q/2)$, with $r_{\rm in}=0.45$,
$r_{\rm out}=1.35$, and $N_k=201$. The sequence is $0\to2\to0$. The two
integer labels agree within $10^{-14}$, and the largest principal
plaquette flux is $2.86$. This flux bound alone does not certify a
continuum sector; Appendix~\ref{app:even-chern-numerics} supplies the
refinement study. Right: the minimum along $k_y=0$, which contains the
relevant primary zeros. The analytic gap closes at each Fermi-sheet
crossing; the sampled points approach but do not land on both closings.
Reversing $\chi_\Delta$ gives the intermediate value $-2$.}
  \label{fig:even-chern-scan}
\end{figure}

The square-lattice annular test is a complementary control. Its gap is
odd under $\bm k\to-\bm k$, as a single spin-polarized pairing
component requires, and its central vortex lies in the inner hole.
Moving the other two zeros across the occupied annulus gives
$C_{\rm band}:0\to2\to0$ (Fig.~\ref{fig:even-chern-scan}). The band
gap closes at each crossing. Reversing the chirality changes the
intermediate value to $-2$. The scan agrees with the oriented vortex
count at all 54 reported radii, and representative sectors are stable
under grid refinement up to $N_k=401$.

The scan realizes the annular vortex rule~\cite{LeNir2026} directly,
and it constrains experiment. Odd pairing need not give an odd Chern
number, because an annulus can exclude the central zero and keep only
a symmetry-related pair. An even thermal plateau therefore cannot, by
itself, identify multivalley pairing. Conversely, a zero plateau
constrains net chiral transport but does not exclude all boundary
states or unresolved domains. Model definitions, the gap-phase plot,
and resolution data are in
Appendices~\ref{subsec:vortex-rule} and~\ref{app:even-chern-numerics}.

\onecolumngrid
\twocolumngrid

\subsection{From the excitation spectrum to the thermal signal}
\label{subsec:envelope}

At finite temperature, quasiparticles sample a range of energies and
Berry curvatures. For a clean quadratic two-band spectrum
$E_\pm=\pm E_{\bm k}$, the transport coefficient, including energy
magnetization and the physical Nambu factor, is
\begin{equation}
  \begin{aligned}
  \frac{\kappa_{xy}}{T}
  &= \frac{\pi^2k_B^2}{6h}\,\frac{1}{2\pi}
     \int \dd^2k\;\Omega_{\bm{k}}\,
     \mathcal{G}\!\left(\tfrac{E_{\bm{k}}}{2k_BT}\right) \\
  &= \frac{\pi^2k_B^2}{6h}\,\CBdG\,
    \Bigl\langle\, \mathcal{G}\!\left(\tfrac{E_{\bm{k}}}{2k_BT}\right)
    \Bigr\rangle_{\!\Omega},\qquad \CBdG\ne0.
  \end{aligned}
\label{eq:kxy-exact-kernel}
\end{equation}
The kernel is
\begin{align}
  \mathcal{G}(x)
  &= \frac{12}{\pi^2}\int_0^{x}\! u^2\,\mathrm{sech}^2 u\;\dd u
\nonumber\\
  &= 1 - \frac{12}{\pi^2}\Bigl[\,x^2\bigl(1-\tanh x\bigr)
     + 2x\ln\!\left(1+e^{-2x}\right)
\nonumber\\
  &\qquad\qquad\quad
     - \,\mathrm{Li}_2\!\left(-e^{-2x}\right)\Bigr],
\label{eq:kernel-G}
\end{align}
and the signed average is
\begin{equation}
  \langle X \rangle_\Omega \equiv
  \frac{\int \dd^2k\;\Omega_{\bm{k}}\,X(\bm{k})}
       {\int \dd^2k\;\Omega_{\bm{k}}}
\label{eq:signed-average}
\end{equation}
This average exists only when $\CBdG\ne0$. The integral in the first
line of Eq.~\eqref{eq:kxy-exact-kernel} also applies when the Chern
number vanishes. Because $\Omega_{\bm k}$ can change sign, the response
need not fall monotonically or stay between zero and its plateau. The
derivation follows the band transport formalism of
Refs.~\cite{QinNiuShi2011,SumiyoshiFujimoto2013} and is given in
Appendix~\ref{app:exact-kernel}.

For the finite-momentum spectrum, $\mathcal G$ is replaced by
\begin{equation}
  \mathcal{W}_{\bm{Q}}(\bm{k})=
  \frac{1}{2}\left[
  \mathcal{G}\!\left(\frac{\eta+\xi_a}{2k_BT}\right)+
  \mathcal{G}\!\left(\frac{\eta-\xi_a}{2k_BT}\right)\right],
\label{eq:branch-weight}
\end{equation}
where $\mathcal G$ is extended as an odd function. This expression
remains valid when the quadratic model is gapless. Its zero-temperature
limit expresses the loss of quantization directly:
\begin{equation}
 \frac{\kappa_{xy}}{Tq_M}\bigg|_{T\to0}
 = C_{\rm band}-\frac{1}{2\pi}
 \int_{|\xi_a|>\eta}\!\dd^2k\,\Omega_{\bm k}.
 \label{eq:curvature-blocking}
\end{equation}
The formula assumes isolated bands, $\eta>0$, and zero-energy contours
of zero area. Where both bands are filled, their opposite curvatures
cancel. Where both are empty, nothing contributes. The loss of
quantization is thus the Berry curvature removed by the pockets,
consistent with Ref.~\cite{Zeng2026}. The deviation is signed and is
not fixed by the pocket area alone.

The same kernel gives a conservative error bound when the bulk is
gapped. Let $A_\Omega=(2\pi)^{-1}\int\dd^2k\,|\Omega_{\bm k}|$.
Monotonicity of $\mathcal G$ and $\eta\pm\xi_a\geq\delta_{\rm BdG}$ give
\begin{equation}
 \left|\frac{\kappa_{xy}}{Tq_M}-\CBdG\right|
 \leq A_\Omega\left[1-\mathcal G\!\left(
 \frac{\delta_{\rm BdG}}{2k_BT}\right)\right].
 \label{eq:plateau-error-bound}
\end{equation}
For a target fractional accuracy $\epsilon$ and $\CBdG\ne0$, it is
enough that the right side be smaller than $\epsilon|\CBdG|$. The bound
turns a measured or calculated gap, together with the absolute
curvature weight, into a temperature window for resolving an integer
plateau.

The grey $\tanh$ curves in Fig.~\ref{fig:fig1_schematic} are a
single-mass comparison. Their relation to the parity-odd determinant,
and their difference from the thermal transport kernel, are derived in
Appendices~\ref{subsec:kernel}--\ref{app:protection-detail}.

\subsection{Disorder and the mobility gap}
\label{subsec:disorder}

The band invariant, the gap criterion, and the kernel describe a clean
crystal. A real device also contains disorder. Disorder can localize
low-energy states, but it can also open new paths for heat through the
bulk. We therefore build the warped finite-momentum model in real
space and compute its heat transport by scattering theory with
Kwant~\cite{Groth2014} (Methods and Appendix~\ref{app:kwant}).
A lattice-resolved parameter set keeps $C_{\rm band}=2$ and the
gapped-to-pocketed sequence of Fig.~\ref{fig:warped}c. It also places
the Fermi wavelength, the coherence length, and the pockets on lattice
scales that simulations can reach. We compare a gapped point G, with
$\delta_{\rm BdG}=0.163$, with four pocketed points P1--P4 whose pockets
grow beyond the onset $w_c=2.09$. The on-site disorder is uniform in
$[-U,U]$ and reaches $U=3$, beyond the Fermi energy $E_F=0.66$.
We express heat transport in Majorana units, $g=K/(q_MT)$, on square
cylinders without edges. We read the integer independently from the
winding of the reflection determinant under a flux
twist~\cite{Fulga2012}, and from the Bott
index~\cite{LoringHastings2010}.

\begin{figure*}[t]
  \setlength{\columnwidth}{\textwidth}
  \setlength{\hsize}{\textwidth}
  \setlength{\linewidth}{\textwidth}
  \centering
  \includegraphics[width=0.86\textwidth]{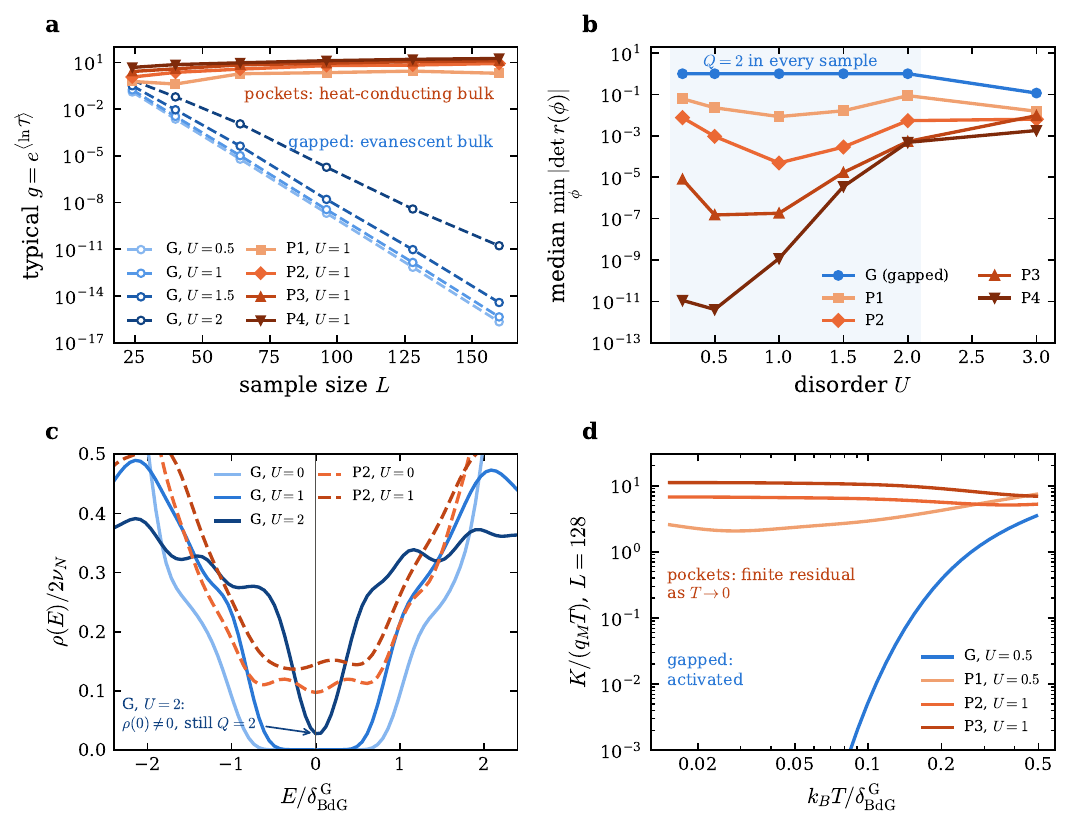}
  \caption{Disorder test of the transport window in the warped
finite-momentum model. G is gapped ($w=1$); P1--P4 ($w=2.2$, $2.5$,
$3$, $4$) have Bogoliubov pockets of growing size. All have
$C_{\rm band}=2$. (a) Typical bulk heat conductance
$g=\exp\langle\ln\mathcal T\rangle$ of square cylinders in Majorana
units. Dashed: G at four disorder strengths, decaying exponentially.
Solid: pocketed points at $U=1$, conducting. (b) Median over samples
of $\min_\phi|\det r(\phi)|$ during a flux cycle at $L=128$. At G it
stays at unity, and the scattering Chern number is $Q=2$ in every
sample up to $U=2$ (shaded). For pockets it nearly vanishes and $Q$ is
not quantized. (c) Kernel-polynomial density of states, normalized by
the normal-state value $2\nu_N$. At $U=2$ localized states fill the gap
of G while $Q=2$ persists. Pockets give a finite $\rho(0)$ even without
disorder. (d) Heat conductance at $L=128$ from energy-resolved
transmission: activated at G and finite as $T\to0$ for pockets.
Energies are in units of the clean gap of G,
$\delta^{\rm G}_{\rm BdG}=0.163$. Averages use 24 disorder realizations
in (a), 12 in (b), and 8 in (d); (c) uses one realization on a torus of
$96$ lattice units.}
  \label{fig:disorder}
\end{figure*}

Figure~\ref{fig:disorder} shows two sharply different responses.
At G the bulk heat transmission decays exponentially with sample size.
Its decay length is $4.0$ lattice units up to $U=1$ and grows only to
$5.4$ at $U=2$. All 12 samples at every disorder strength up to $U=2$
return $Q=2$, with $\min_\phi|\det r|\geq0.98$, and the Bott index
equals two on every torus. The integer therefore survives disorder
whose root-mean-square strength is seven times the excitation gap. It
is lost only at $U=3$, more than four times $E_F$. At $U=2$ disorder
states fill the gap, so $\rho(0)\neq0$, yet the integer remains because
these states are localized. In a disordered sample the plateau
therefore requires a mobility gap. The spectral gap of
Eq.~\eqref{eq:gapped-condition} is its clean limit.

The pocketed points respond in the opposite way, at the same
$C_{\rm band}$ and the same disorder. Their bulk conducts heat: $g$
grows with sample size and reaches $2$--$18$ Majorana quanta at $L=160$
and $U=1$. The reflection determinant nearly vanishes during the flux
cycle, with medians from about $10^{-1}$ down to $10^{-11}$, so the
winding no longer defines an integer. The zero-energy density of states
is finite, and disorder does not localize the pocket states at any
strength studied. The thermal signatures differ accordingly
(Fig.~\ref{fig:disorder}d). The gapped bulk freezes out with
activation. A pocketed bulk keeps a residual $K/T$ of $2$ to $11$
Majorana quanta as $T\to0$, which grows on cooling for the larger
pockets. At the strongest disorder, the pocketed conductance grows
logarithmically with sample size. The slopes, $0.23$--$0.28$ for P1--P3,
match a calibrated class-D thermal metal, with $0.26\pm0.09$ and
$0.28\pm0.08$~\cite{Chalker2001,Zakharov2025}. Extended Bogoliubov
pockets, rather than a small spectral gap, are what remove the
quantized readout.

\subsection{A quantitative domain-wall prediction}
\label{sec:predictions}
\paragraph{Reverse the domain and measure the sign.}
\label{subsec:cosign_test}

A plateau measures the total chiral invariant. Domain control tests
whether its orientation is inherited from the parent magnetization.
For two time-reversed but otherwise equivalent superconducting states,
$\CBdG$ changes sign. In a chirality-inheritance scenario such as
Ref.~\cite{li2025berrytrashcan}, the anomalous Hall signal and the
thermal Hall signal should both reverse. With a fixed laboratory
convention, the proposed correlation reads
\begin{equation}
  \sgn\!\left[\frac{\kappa_{xy}}{T}\right]_{\rm domain}
  =\sgn[R_{xy}]_{\rm domain},
\label{eq:cosign}
\end{equation}
The relative sign must be calibrated. Its absolute value is not a
universal property of all microscopic pairing states.

The protocol is direct. Prepare a nearly uniform domain, measure its
low-temperature thermal Hall response, reverse the domain, and repeat
at the same density and field. Imaging the domains before and after
the thermal measurement checks that heating or the probe current has
not moved the walls. This control matters because the domains respond
to very small currents~\cite{Dutta2026}. A failure of correlated
reversal would challenge simple chirality inheritance. A changed
magnitude would call for checks of domain fractions, gap size, and
contact equilibration before a change of topological sector is
assigned.

\paragraph{Write a domain wall and count its heat channels.}
\label{subsec:domain-wall}

The most direct quantitative prediction links the two bulk plateaux to
the heat carried along their interface. For fully gapped domains with
$C_L=+C$ and $C_R=-C$, bulk--boundary correspondence requires a net
number
\begin{equation}
  |\Delta C|=|C_L-C_R|=2|C|
\label{eq:domain-deltaC}
\end{equation}
of chiral Majorana modes. If these protected channels dominate
transport and equilibrate with ideal reservoirs, the low-temperature
wall conductance obeys
\begin{equation}
 \boxed{\begin{aligned}
  \lim_{T\to0}\frac{K_{\rm DW}}{T}
  &=\left|\lim_{T\to0}\frac{\kappa_{xy}^{L}-\kappa_{xy}^{R}}{T}\right|\\
  &=2|C|\frac{\pi^2k_B^2}{6h}
   =|C|\frac{\pi^2k_B^2}{3h}.
 \end{aligned}}
\label{eq:domain-wall-K}
\end{equation}
The wall thus carries twice the magnitude of either bulk plateau.
For $|C|=1$, the two bulk values are
$\kappa_{xy}/T=\pm4.732\times10^{-13}\,\mathrm{W\,K^{-2}}$,
while $K_{\rm DW}/T=9.464\times10^{-13}\,\mathrm{W\,K^{-2}}$.
Two Majorana channels then carry one Dirac thermal quantum.

This relation suggests a reversible experiment. First establish the
opposite bulk plateaux. Then write a wall between the domains, and use
local heat injection and directional thermometry to resolve transport
along it. Finally move or erase the wall and repeat with the same
calibrated contacts. A heat path that follows the wall and satisfies
Eq.~\eqref{eq:domain-wall-K} would test the same integer in the bulk
and at its boundary. Together with correlated domain reversal, it
would also connect superconducting chirality to the magnetic domain
pattern.

Here $K_{\rm DW}$ denotes the conductance of the wall channels.
A measured device conductance also depends on the contact geometry,
other edges, and parasitic heat paths. A broad normal strip or
additional gapless modes can spoil the predicted quantization. The
resistive walls observed in Ref.~\cite{Dutta2026} offer a controllable
setting for this test; their heat-channel content remains to be
measured.

\begin{figure*}[t]
  \setlength{\columnwidth}{\textwidth}
  \setlength{\hsize}{\textwidth}
  \setlength{\linewidth}{\textwidth}
  \centering
  \includegraphics[width=0.86\textwidth]{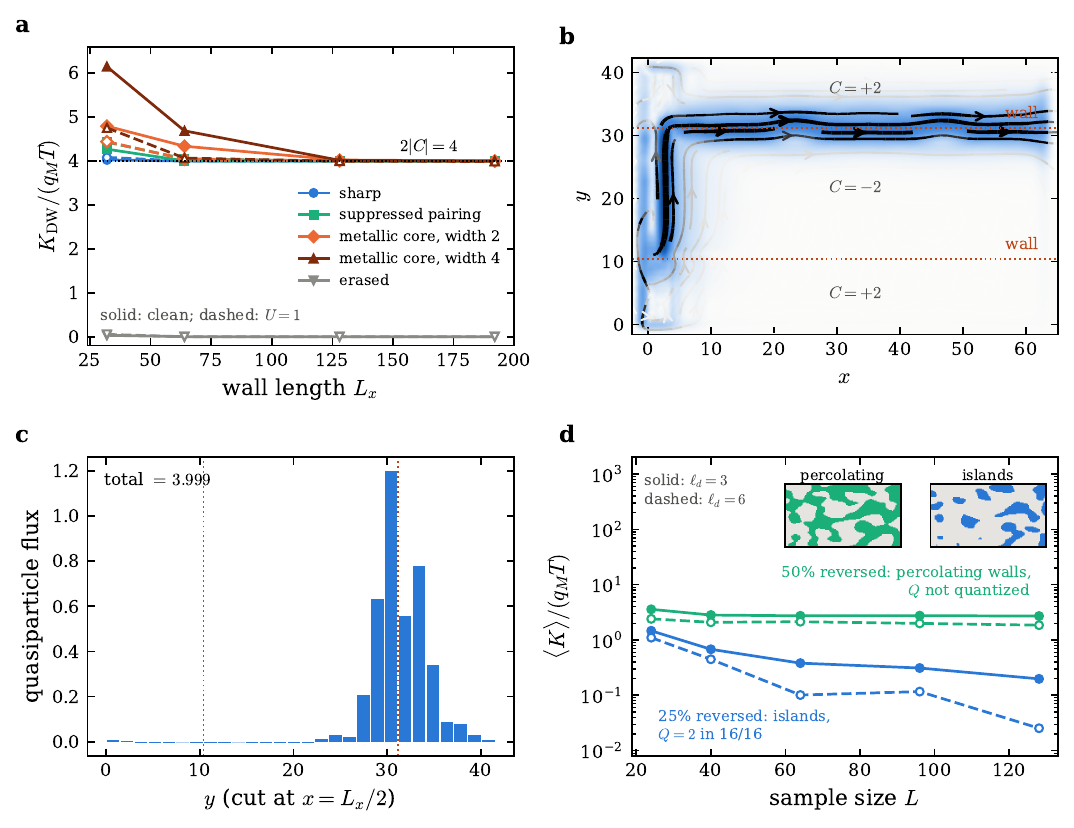}
  \caption{Written walls and domain networks at the gapped point G.
(a) Heat conductance of a cylinder with two walls, in Majorana units,
against wall length. Sharp walls give $2|C|=4$. Walls with suppressed
pairing (width 3, pairing reduced to $0.3$) or a metallic core (width
2 or 4) converge to the same value as their extra channels are
filtered. Erased walls give zero. Solid: clean; dashed: $U=1$, eight
realizations. (b) Quasiparticle current injected from the left lead at
$E=0.02$. Heat enters along the contact and follows the wall whose
Majorana modes move to the right. Dotted lines mark the walls, and the
labels give the domain Chern numbers. (c) Flux through the cut
$x=L_x/2$, summing to $3.999$. (d) Random textures with correlation
length $\ell_d$, averaged over 16 textures. Reversed islands (25\%)
keep the bulk insulating with $Q=2$. Percolating domains (50\%) conduct
heat. Insets: representative textures.}
  \label{fig:walls}
\end{figure*}

\paragraph{Microscopic test of written walls.}
The time-reversed domain is the complex conjugate of the BdG
Hamiltonian, with opposite valley, chirality, and pair momentum.
Writing two such walls into a gapped cylinder switches its heat
transmission from zero to $4.000=2|C|$ (Fig.~\ref{fig:walls}a). The
deviation is within $10^{-4}$ for walls of $64$ lattice units and
longer, and within $2\times10^{-3}$ at disorder $U=1$. Erasing the
walls returns the transmission to zero. The heat follows the wall whose
modes move downstream (Fig.~\ref{fig:walls}b), and the flux through a
cut across the sample is $3.999$ quanta (Fig.~\ref{fig:walls}c).

Real walls need not be sharp or fully gapped. We therefore also study
walls with suppressed pairing and walls with a metallic core. These
first carry additional nonchiral channels, up to $6.15$ quanta for a
core four lattice units wide. The extra channels decay exponentially
with wall length, over $8$--$34$ lattice units, and every wall type
converges to $2|C|$ within $0.01$. Disorder shortens this approach.
A quantized wall therefore requires neither an atomically sharp nor a
fully gapped interface. This matters for valley domain walls, which
can host intervalley structure and block charge across
them~\cite{Phong2026}. Our results suggest that such walls can still
guide quantized heat along their length.

Random written textures extend the test to networks
(Fig.~\ref{fig:walls}d). Isolated reversed islands covering a quarter
of the sample leave the bulk insulating. They also leave the integer of
the majority domain intact, with $Q=2$ in all 16 samples. When the
reversed domains percolate, their walls form a network that carries
about two to three quanta through the bulk, nearly independent of
size. The integer is then no longer quantized. A plateau measurement
therefore reads the domain that percolates. Writing domains can switch
the bulk between a heat insulator and a heat conductor.

\paragraph{Establish the transport regime.}

Noise thermometry has resolved thermal quanta in quantum Hall devices
and in encapsulated graphene~\cite{Banerjee2018,Srivastav2019,Srivastav2022}.
Nonlocal thermal measurements also work in van der Waals
structures~\cite{Waissman2021}. These techniques offer a route to the
required sensitivity. Superconducting devices still need independent
checks of energy transfer into neutral edge modes, phonon leakage, and
contact equilibration.

A floating contact measures outgoing heat conductance. On its own it
does not determine the sign of $\kappa_{xy}$. Directional,
multiterminal thermometry or an equivalent calibrated Hall geometry is
needed for the domain-reversal test. Longitudinal electronic heat
transport and gap spectroscopy should be measured alongside it.
Activated bulk transport supports the gapped interpretation. In the
real-space calculations of Fig.~\ref{fig:disorder}d, a pocketed bulk
instead keeps a residual $K/T$ of several Majorana quanta as $T\to0$.
A residual linear term warns of low-energy excitations, but is not
specific to a Bogoliubov Fermi surface. Localization can also suppress
longitudinal transport in a system without a spectral gap.

\section{Discussion}
\label{sec:conclusion}

Our calculations connect three ingredients that decide whether heat
transport can read superconducting topology: band topology, the
occupation of quasiparticle states, and the thermal weighting of their
Berry curvature. Separating them has measurable consequences. Warping
can keep an isolated-band Chern number while creating zero-energy
pockets that remove the thermal plateau. The exact kernel quantifies
this loss and gives a sufficient criterion for plateau accuracy when
the bulk stays gapped. Real-space calculations with disorder sharpen
the picture. The integer survives as long as a mobility gap remains,
even when localized states fill the spectral gap. The extended states
of Bogoliubov pockets remove it.

The annular result changes how a plateau should be read. An even
integer can arise within a single odd pairing component. The measured
integer therefore fixes the net chiral Majorana content under the
stated conditions, while leaving room for different valley and pocket
structures. This makes the readout useful when quantum oscillations do
not fix a unique Fermi surface.

For a gapped chiral state, the results translate into four measurable
predictions.
\begin{enumerate}
\item The low-temperature plateau equals $\CBdG q_M$, with
$q_M=4.732\times10^{-13}\,\mathrm{W\,K^{-2}}$, and it reverses with the
domain together with the anomalous Hall signal
[Eqs.~\eqref{eq:kxy-intro-quantum} and~\eqref{eq:cosign}].
\item A written wall between time-reversed domains carries
$K_{\rm DW}/T=2|\CBdG|q_M$, and erasing the wall removes this channel
[Eq.~\eqref{eq:domain-wall-K} and Fig.~\ref{fig:walls}a].
\item A gapped bulk freezes out with activation. A pocketed bulk keeps a
residual $K/T$ as $T\to0$ and an unquantized transverse response
(Fig.~\ref{fig:disorder}d).
\item With imaged domains, the plateau follows the domain that
percolates across the sample. Isolated reversed islands leave it
unchanged, while a percolating network of walls conducts heat through
the bulk (Fig.~\ref{fig:walls}d).
\end{enumerate}

The model establishes a mechanism and a measurement protocol.
Material-specific thresholds require a self-consistent multiband
calculation with the parent-band embedding. Quantitative device signals
also require a model of the contacts and the domain network. Above
$T_c$, the continuation in Fig.~\ref{fig:fig1_schematic} remains
conditional on long-lived chiral correlations and a consistent
transport theory.

The immediate opportunity is to compare a bulk plateau, its reversal
with an imaged domain, and a written wall whose conductance matches
their difference. Our microscopic calculations support this test.
A written wall carries $2|C|$ Majorana channels even with a metallic
core, and only a percolating network of walls makes the bulk conduct
heat. Gate sweeps can add a further test by linking a change of the
integer to a gap closing. Observing the linked bulk and wall response
would turn domain control into a quantitative probe of superconducting
topology. It would also show that protected heat flow can be routed
through a rewritable domain pattern.

\section{Methods}
\label{sec:methods}
\subsection{Normal-state model and conventions}
\label{subsec:dispersion}

For reference, the isotropic low-energy dispersion
is~\cite{Ghazaryan2023,Geier2025}
\begin{equation}
  \varepsilon_k = D\sqrt{1 + (k/k_0)^{2N}} + \frac{\hbar^2 k^2}{2m},
\label{eq:dispersion}
\end{equation}
Here $D$ is an interlayer energy scale, $N$ the layer number, and $k_0$
and $m$ are effective parameters. For $N>1$, a negative quadratic
coefficient, $m<0$, permits a Mexican-hat minimum; large $D$ alone does
not imply it. The layer Berry phase and remote hopping structure are
discussed in Refs.~\cite{Koshino2009,Slizovskiy2019}. The circular and
annular dispersions used here are benchmarks. They do not fit the
multitone normal state inside the measured dome.

At zero additional pair momentum the Nambu matrix is
\begin{equation}
  H_{\rm BdG}(\bm{k})
  = \begin{pmatrix}
      \xi_{\bm{k}} & \Delta_{\bm{k}} \\
      \Delta_{\bm{k}}^* & -\xi_{-\bm{k}}
    \end{pmatrix},
\label{eq:BdG-ham}
\end{equation}
with $\xi_{\bm k}=\varepsilon_{\bm k}-\mu$ and a chiral gap
$\Delta_{\bm k}=\Delta_0(k_x+i\tau k_y)/k_F$ near a valley.
The sign assigned to this gap depends on the Berry and coordinate
conventions. We fix the numerical convention by the square-lattice
anchor
\begin{equation}
  H_{\rm BdG}(\bm{k}) = d_x\tau_x + d_y\tau_y + d_z\tau_z,
\label{eq:pwave-bdg}
\end{equation}
with $d_x=\Delta_0\sin k_x$, $d_y=\Delta_0\sin k_y$, and
$d_z=2t(2-\cos k_x-\cos k_y)-\mu$. The FHS link orientation used here
gives $\CBdG=+1$ for $0<\mu<4t$, so the corresponding off-diagonal gap
is $\Delta_0(\sin k_x-i\sin k_y)$. In this convention, the oriented
charge in the vortex rule is minus the counterclockwise winding of
$\arg\Delta$. This distinction fixes the signs in the annular tests
and in Appendix~\ref{app:read-green}.

\subsection{Numerical protocols and scope}
\label{subsec:numerics}

FHS link variables~\cite{Fukui2005} give the Chern number of the lower
band. We report the largest principal plaquette flux and refine the
mesh near sector boundaries. A principal argument already lies in
$[-\pi,\pi]$, so a value below $\pi$ does not by itself certify the
continuum sector. Agreement under refinement and analytic gap-closing
conditions provide the relevant checks. A roundoff-level deviation from
an integer measures discrete integrality, not physical accuracy.

The warped map uses $241\times241$ points for the Chern number and
$481\times481$ for the gap. The annular refinement reaches
$401\times401$. Gap values quoted from finite grids are sampled
estimates. The known vortex crossings also give exact conditions for a
closing. Continuous minimization on the reference slice checks the sign
of small grid minima. The scan fixes the pair momentum along $x$ and
holds the pairing texture fixed. It does not search over competing
orders or momentum directions.

The supporting calculations include Wilson-loop flux threading and
28 DMRG runs on $L_x=8$ cylinders. The latter use TeNPy,
$\mathbb Z_2$ parity conservation, and a bond-dimension ramp from
64 to 512~\cite{Hauschild2024}. The interaction is $V_{nn}=0$ and the
complex pairing field is explicit. These runs benchmark a quadratic
finite-cylinder ground state. The reported Chern numbers come from the
Bloch Hamiltonian, not from a many-body invariant of the MPS.

Real-space transport calculations use Kwant~\cite{Groth2014}. The
Bloch functions of the warped model are inverted exactly into lattice
hoppings, with ranges of two and three lattice vectors for the normal
and pairing terms. Tori use a gauge frame with uniform pairing. Leads
and walls use the laboratory frame with pair phase
$e^{i\bm Q\cdot\bm R}$. The two frames agree to $2\times10^{-15}$, and
torus spectra match the Bloch spectra to $10^{-13}$. The heat
conductance in Majorana units is the Nambu-trace transmission between
normal-metal leads at $E\to0$, so one chiral Majorana mode contributes
one. The scattering Chern number uses leads of decoupled chains and a
twist acting on the full Nambu spinor~\cite{Fulga2012}. Densities of
states and the real-space Kubo--Bastin Hall function use the kernel
polynomial method~\cite{Weisse2006,Garcia2015}. Benchmarks give the
edge transmission $2.001=|C|$, a two-wall transmission of $3.999$,
$Q=\pm2$ under chirality reversal, a Bott index of two, and a
real-space Hall function $s(0)=2.013\pm0.031$ at G. The parameters,
ensembles, and complete data are given in Appendix~\ref{app:kwant}.

Details, all numerical plots, and per-parameter tables follow in the
appendices. The accompanying code and summary data specify the
numerical models, observables, and independent quadratic benchmarks.

\renewcommand{\bibfont}{\fontsize{9}{10}\selectfont}
\bibliographystyle{apsrev4-2}
\bibliography{ref_rhg_v18}

\clearpage
\appendix
\onecolumngrid
\raggedbottom

\section{Vortex rule and annular model}
\label{subsec:vortex-rule}

The occupied-vortex formula of Ref.~\cite{LeNir2026} is
\begin{equation}
  \CBdG = \mathcal{Q}
        = \sum_{i:\,\bm{k}_i \in D_{\rm occ}} \ell_i,
\label{eq:CBdG-vortex}
\end{equation}
For disconnected pairing-relevant regions, $D_s=\bigcup_aD_a$, it reads
\begin{equation}
  \CBdG=\sum_a\sum_{i:\,\bm{k}_i\in D_a}\ell_i,
\label{eq:CBdG-vortex-multitone}
\end{equation}
The charges $\ell_i$ use the same orientation as $w_i$ in the main text.
In the gapped two-band model, $D_{\rm occ}=D_s$. At finite pair momentum
it is $\xi_s$, rather than the unsymmetrized normal dispersion, that
sets this region. In a projected physical Chern band the pairing
amplitude is gauge dependent. A smooth local gauge, its transition
functions, and the parent-band embedding must be specified. Remote
occupied sectors can contribute to the total physical invariant.
Our periodic proxy has a globally defined orbital basis, so this
additional material calculation is absent rather than implicit.

\subsection{Annular test geometry}
\label{subsec:annulus}

The square-lattice test uses
\begin{align}
  \xi_{\rm ann}(\bm{k}) &=
  \bigl(\rho^2(\bm{k}) - r_{\rm in}^2\bigr)
  \bigl(\rho^2(\bm{k}) - r_{\rm out}^2\bigr),
    \nonumber \\
  \rho^2 &= 4 - 2\cos k_x - 2\cos k_y,
\label{eq:xi-ann}
\end{align}
with $r_{\rm in}=0.45$ and $r_{\rm out}=1.35$. The region
$r_{\rm in}<\rho<r_{\rm out}$ is occupied. We impose
\begin{align}
  \Delta(\bm{k}) = \Delta_0\,&
    (\sin k_x + i\chi_\Delta\sin k_y) \nonumber\\
    &\times (\sin k_x - \sin q + i\chi_\Delta\sin k_y) \nonumber\\
    &\times (\sin k_x + \sin q + i\chi_\Delta\sin k_y),
\label{eq:gap-finite-vortex}
\end{align}
with $\Delta_0=0.50$ and $\chi_\Delta=-1$. Near the zone centre, the
three zeros are at $(0,0)$ and $(\pm q,0)$. Their oriented charges are
$+1$, although the counterclockwise phase winding of the off-diagonal
gap is $-1$ for this sign of $\chi_\Delta$.
The polynomial is odd: $\Delta(-\bm k)=-\Delta(\bm k)$.
It therefore supplies an explicit single-component annular example
with even Chern number. Periodic image zeros lie outside the occupied
region over the reported scan. They are discussed in
Appendix~\ref{app:even-chern-numerics}.

\begin{figure}[H]
  \centering
  \includegraphics[width=0.6\columnwidth]{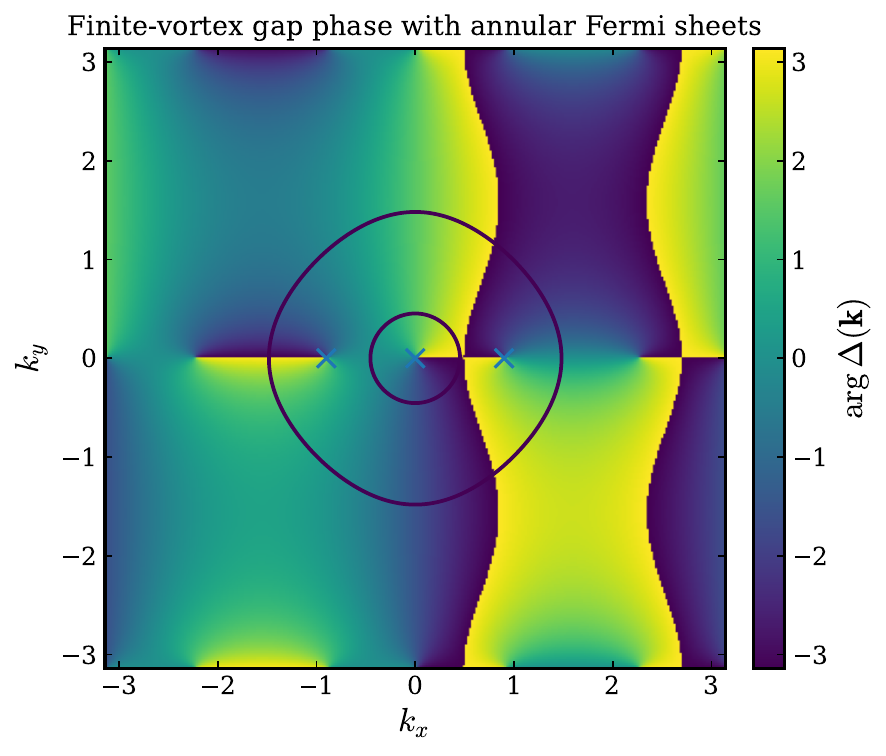}
  \caption{Gap phase in the annular test. Solid contours are the two Fermi sheets.
The central zero lies in the inner hole; the finite-momentum pair lies
inside the occupied annulus. For $\chi_\Delta=-1$, each primary zero
has counterclockwise gap-phase winding $-1$ and oriented Chern charge
$+1$ in the numerical convention. The occupied charge is two, matching
the computed band invariant. The displayed texture is imposed rather
than obtained from a gap equation.}
  \label{fig:gap-phase}
\end{figure}

The vortex texture is imposed. It tests how a specified gap texture
maps to topology, rather than demonstrating self-consistent vortex
nucleation. The latter is treated in
Refs.~\cite{LeNir2026,Patri2025}. The relevant control parameter for the
square-lattice scan is the primary pair radius $2\sin(q/2)$.
The minimum along $k_y=0$ approaches zero at the two analytic crossing
radii. Figure~\ref{fig:gap-phase} shows the gap-phase texture and its
position relative to the two Fermi sheets. Grid and chirality checks
are given in Appendix~\ref{app:even-chern-numerics}.

\section{Pairing scenarios and absolute thermal scales}
\label{subsec:chern-scenarios}

\paragraph{Single-component disk.}
A simply connected region that contains one oriented unit vortex gives
\begin{equation}
  \CBdG=1,
\label{eq:C-pip}
\end{equation}
This is the weak-pairing anchor with one net chiral Majorana mode.
Short-range attraction can select an orientation tied to the parent
Berry curvature~\cite{li2025berrytrashcan}. Other interactions and
nonuniform curvature can favour additional vortices and higher
integers~\cite{Patri2025,MayMann2026}. These are microscopic selection
mechanisms, rather than different thermal normalizations.

\paragraph{Annular and multicomponent states.}
A minimal central vortex contributes zero when it lies in an unoccupied
inner hole~\cite{Geier2025}. A finite-momentum pair inside an annulus
instead gives $\CBdG=2$, even for one odd pairing component. The odd
angular-momentum restriction for spin-polarized intravalley pairing does
not imply odd Chern parity for every Fermi-surface topology.
An even value therefore does not exclude the single-component class.

Several components can also produce an even total. A chiral
$d_{x^2-y^2}+id_{xy}$ gap has phase winding two in a suitable block.
Its physical thermal conductance requires the Chern number of the full
BdG Hamiltonian, including the independent spin and valley sectors.
A block label must not be substituted for this total. A measured
$\CBdG=2$ is consequently a two-Majorana benchmark, not a unique
identification of $d+id$ pairing.

\paragraph{Other Chern bands.}
For the projected $(1{+}n)$ structures considered in
Ref.~\cite{LeNir2026}, a parent band with $C_P=n$ constrains the total
gap vorticity to $V_\Delta=2C_P$. The occupied share can reach
$\CBdG=2n-1$ in the simply connected examples studied there.
This is a possible family of states, rather than a value fixed by
layer number alone.

\subsection{Plateau heights}
\label{subsec:prefactor}
\begin{align}
  \CBdG = 1\ (p+ip): \quad
  \frac{\kappa_{xy}}{T}\Big|_{T\to0}
    &= 4.732\times10^{-13}\;\frac{\mathrm{W}}{\mathrm{K}^2},
\label{eq:pred-C1}\\[2pt]
  \CBdG = 2\ (\text{total}): \quad
  \frac{\kappa_{xy}}{T}\Big|_{T\to0}
    &= 9.464\times10^{-13}\;\frac{\mathrm{W}}{\mathrm{K}^2}.
\label{eq:pred-C2}
\end{align}
The first line is the one-mode anchor and the second a total-$C=2$
benchmark. Their ratio is independent of gap magnitude in the fully
gapped low-temperature limit. At finite temperature the appropriate
scale is the minimum excitation gap at that temperature, not
$\Delta_0$ alone. A large zero-temperature ratio
$2\Delta_0/(k_BT_c)$ does not ensure a plateau near $T_c$ if the gap
closes there. Equation~\eqref{eq:plateau-error-bound} gives a sufficient
criterion when the two-band description applies.

\subsection{Reading an integer}
\label{subsec:discriminator}
\begin{equation}
  \frac{\kappa_{xy}}{T}\bigg|_{T \to 0}
  = \CBdG \cdot \frac{\pi^2 k_B^2}{6h}
  \approx \CBdG \times 4.73 \times 10^{-13}\,
    \frac{\text{W}}{\text{K}^2}.
\label{eq:kxy-sat}
\end{equation}
For candidate totals $1,2,3,5$, the magnitudes are
\begin{align}
  \frac{\kappa_{xy}}{T}\bigg|_{T\to 0}
  &= (4.73,\;9.46,\;14.2,\;23.7)\times10^{-13}
     \frac{\rm W}{\rm K^2},\notag\\
  &\hspace{2.5em}\text{for }\CBdG=(1,2,3,5).
\label{eq:staircase-C}
\end{align}
These values distinguish topological sectors. They do not uniquely
identify their microscopic origin. A zero result constrains the net
chiral response only after domain cancellation, thermal leakage, and
equilibration have been controlled.

A heated floating contact measures the sum of outgoing thermal channels.
For ideal channels of total coefficient $g=K/T$, the heat balance is
$\Delta P=\tfrac12g(T_m^2-T_0^2)$ before parasitic terms are added.
For an isolated, fully chiral boundary $K/\kappa_0=|\CBdG|/2$, where
$\kappa_0=\pi^2k_B^2T/(3h)$. Directional thermometry is needed to recover
the sign. Existing quantum Hall measurements demonstrate sensitivity
to thermal quanta at millikelvin temperatures
\cite{Banerjee2018,Srivastav2019,Srivastav2022}; nonlocal graphene
thermometry supplies a complementary method~\cite{Waissman2021}.
The cited values of roughly $12$--$20$~mK and percent-scale resolution
are benchmarks from those devices, not demonstrated performance for
superconducting rhombohedral graphene.

In a diffusive Hall geometry, a local tensor model gives
$|\delta T_y|=|\kappa_{yx}^{2\rm D}/\kappa_{yy}^{2\rm D}|
|\nabla_xT|W$ under zero transverse heat flow. Reducing the denominator
to $\kappa_{xx}^{2\rm D}$ requires in-plane isotropy. In an edge-dominated
ballistic device a contact-network description is more appropriate.
Neither expression removes the need to characterize the thermal boundaries.

\section{Numerical benchmarks and their interpretation}
\label{subsec:numerical}

\begin{figure}[H]
  \centering
  \includegraphics[width=0.95\textwidth]{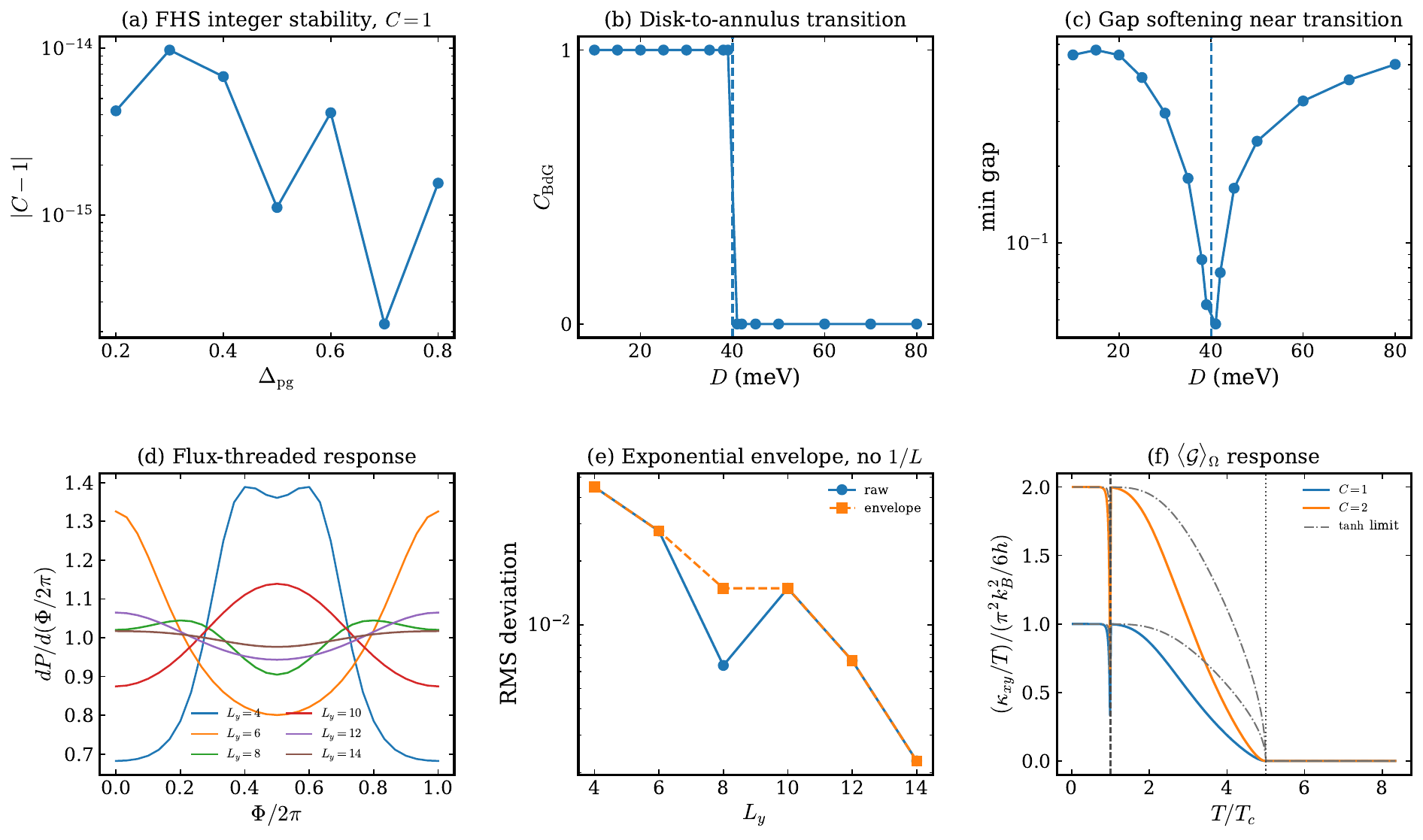}
  \caption{Supporting band and finite-width benchmarks.
(a) Discrete FHS integrality for the square-lattice anchor. The residual
is roundoff, not a measure of continuum accuracy. (b,c) Chern number
and gap in the displacement-field proxy. Its $D_c=40$~meV is an input;
the grid brackets the closing. (d) Flux derivative of the band
polarization for six circumferences. (e) Raw RMS deviations and the
reported exponential fits. The label referring to $c_1=0$ denotes a
comparison with the continuum determinant, not an exclusion of all
algebraic finite-size effects. (f) Supplied illustrative two-gap thermal
curves, with $\Tstar=5T_c$. The BCS and pseudogap branches are stitched
at $T_c$, so the feature there is an ansatz artefact. Above $T_c$ the
curves are conditional quasiparticle illustrations, not a derived total
response. Grey curves are single-mass comparisons. The low-temperature
factor of two represents different total Chern numbers.}
  \label{fig:sixpanel}
\end{figure}

The three numerical approaches answer distinct questions. The FHS
calculation labels the isolated Bloch band. The Wilson-loop calculation
tests flux-dependent band geometry on narrow cylinders. DMRG checks the
quadratic finite-cylinder ground state with explicit pairing.
Figure~\ref{fig:sixpanel} summarizes the band and finite-width
benchmarks. These calculations check distinct parts of the numerical
implementation; they do not establish spontaneous many-body order.

The square-lattice FHS scan returns $\CBdG=1$ throughout the pairing
range, with discrete integrality at roundoff precision. A dispersion
proxy gives a disk-to-annulus transition with $\CBdG:1\to0$ at an
imposed scale $D_c=40$~meV. This scale is a model input. The calculation
does not predict a transition at $40$~meV in a device.
Wilson-loop data at six circumferences show oscillatory finite-width
deviations and admit exponential fits. These short sequences do not
exclude algebraic terms or polynomial prefactors.

The 28 DMRG cases recover the phase relation
\begin{equation}
  \arg\mathcal{A}_y-\arg\mathcal{A}_x=-\frac{\pi}{2}.
\label{eq:pip-signature}
\end{equation}
The largest phase error is approximately $2.8\times10^{-7}$~rad.
The agreement is a check of implementation and convergence because the
Hamiltonian already imposes complex pairing. The energy error per site
ranges from machine precision at $L_y=4$ to $3.7\times10^{-6}$ at
$L_y=10$. Particle--hole residuals and Chern numbers refer to the
matching exact BdG matrices. The MPS data are discussed separately in
Appendix~\ref{app:dmrg}.

\section{Illustrative continuation above the coherent phase}
\label{subsec:two-gap}

A common crossover parametrization separates a condensate amplitude
from a noncondensed-pair spectral scale~\cite{Chen2024RMP}:
\begin{equation}
  \Delta^2(T)=\Dsc^2(T)+\Dpg^2(T),
\label{eq:two-gap}
\end{equation}
This equation describes a spectral ansatz. It does not imply a static
anomalous expectation value when $\Dsc=0$. Nor does a nonzero
$\Dpg$ alone define a BdG Chern number above $T_c$.
A quasiparticle continuation would need long-lived chiral correlations
and a consistent Green-function transport calculation, including its
vertices and collective response.

The single-mass construction defines the comparison envelope
\begin{equation}
  \left.\frac{\kappa_{xy}}{T}\right|_{\rm qp}^{\rm env}
  =\frac{\pi^2k_B^2}{6h}\,\CBdG\,
   \tanh\!\left[\frac{\Dpg(T)}{2k_BT}\right],
\label{eq:kxy-pg}
\end{equation}
The superscript ``env'' identifies an illustrative envelope. It is not
a derived above-$T_c$ response. Replacing it by the band kernel would
still require a justified effective spectrum, its Berry structure,
and sufficiently narrow quasiparticle poles. The scale $\Tstar$ and
the gap ratios in the figures are illustrative inputs.

The experimentally measured coefficient can be organized as
\begin{equation}
 \kappa_{xy}^{\rm meas}
 =\kappa_{xy}^{\rm qp}+\kappa_{xy}^{\rm fl}
  +\kappa_{xy}^{\rm vort}+\kappa_{xy}^{\rm ph},
\label{eq:kxy-decomposition}
\end{equation}
The terms denote quasiparticles, fluctuations, vortices, and phonon or
substrate contributions. Normal-state electronic anomalous transport
must be included in the electronic baseline when pairing is suppressed.
Energy magnetization is already included in the band transport formula;
it is not an additional heat channel. Chiral-fluctuation calculations
of Hall and Nernst effects~\cite{SumiyoshiFujimoto2014} motivate this
separation without supplying a quantitative RHG thermal Hall correction.

\section{Parity-odd determinant and comparison envelope}
\label{subsec:kernel}

The cylinder determinant of
Refs.~\cite{GhoshKlinkhamer2017,Ghosh2026,Ghosh2026BEC} provides a
useful comparison with the thermal kernel. For a two-component massive
Dirac fermion on $\mathbb R_\tau\times\mathbb R_x\times S_L^1$, let
$\psi_v(y+L)=e^{i\alpha_v}\psi_v(y)$ and
$\vartheta_v=\alpha_v+q_v\oint a_y\dd y$. In units $\hbar=k_B=1$, the
mass-dependent parity-odd polarization is
\begin{equation}
  \Pi_{v,\text{odd}}^{\mu\nu,\text{IR}}(p)
  = \frac{iq_v^2}{2\pi}\,
    \mathcal{K}_v^{\text{IR}}(p;L,\vth_v)\,
    \epsilon^{\mu\nu\rho}p_\rho,
\label{eq:Pi-odd}
\end{equation}
with
\begin{equation}
  \boxed{
  \begin{aligned}
  \mathcal{K}_v^{\text{IR}}(p;L,\vth_v)
  &= \frac{\chi_v m_v}{2}\int_0^1 \frac{\dd u}{\Delta_{v,u}}\,\\
  &\quad \times \frac{\sinh(\Leff\Delta_{v,u})}
         {\cosh(\Leff\Delta_{v,u}) - \cos(\vth_v + 2\pi r u)},
  \end{aligned}
  }
\label{eq:K-full}
\end{equation}
Here $\Delta_{v,u}=\sqrt{m_v^2+u(1-u)\tilde p^2}$,
$L_{\rm eff}=L/|v_F|$, $r$ labels the external compact harmonic, and
$\chi_v$ fixes the cone orientation. This is the one-loop infrared
contribution. A gauge-invariant lattice completion also fixes the
mass-independent regulator term. It must not be inferred from the
infrared cone alone.

For a static Euclidean coefficient, compactifying time with
antiperiodic boundary conditions gives
\begin{equation}
  R\!\left(\frac{|m_v|}{T},\pi\right)
  = \frac{\sinh(|m_v|/T)}{\cosh(|m_v|/T) + 1}
  = \tanh\!\left(\frac{|m_v|}{2T}\right),
\label{eq:R-thermal}
\end{equation}
This identity concerns the single-mass static determinant. At finite
temperature, static and transport limits need not commute.
It does not equate an electromagnetic polarization with thermal transport.

We retain the following normalization solely to define the comparison
envelope:
\begin{equation}
  \Kem(T) \equiv 2\CBdG\,\tanh\!\left(\frac{\Delta}{2T}\right).
\label{eq:Kem-T}
\end{equation}
The symbol $K_{\rm env}$ denotes a chosen comparison normalization, rather than a physical electromagnetic Chern--Simons level. The two BdG bands do not supply two independent
identical charged Dirac species. In this notation,
\begin{equation}
  \left(\frac{\kappa_{xy}}{T}\right)_{\!\rm env}
  \equiv \frac{\pi^2 k_B^2}{12h}\,\Kem,
\label{eq:kxy-Kem}
\end{equation}
and hence
\begin{equation}
  \left(\frac{\kappa_{xy}}{T}\right)_{\!\rm env}
  = \frac{\pi^2 k_B^2}{6h}\,\CBdG\,
    \tanh\!\left(\frac{\Delta}{2T}\right).
\label{eq:kxy-T}
\end{equation}
These are envelope definitions. They agree with the physical thermal
response at $T\to0$ in a fully gapped phase, but need not agree at
finite temperature.

The physical class-D normalization follows independently from
$c_-=\CBdG/2$:
\begin{equation}
  \frac{\kappa_{xy}}{T}\bigg|_{T\to0}
  = \CBdG\,\frac{\pi^2k_B^2}{6h}
  \approx \CBdG \times 4.732\times10^{-13}\;
  \frac{\mathrm{W}}{\mathrm{K}^2}.
\label{eq:kxy-majorana-quantum}
\end{equation}
The Dirac quantum is twice this value. The half-integer thermal
conductance observed at filling $5/2$~\cite{Banerjee2018} is a
benchmark for resolving neutral chiral heat transport, not a
measurement of a superconducting BdG Chern number.

\subsection{Asymptotic comparison}
\begin{equation}
  \mathcal{G}(x) \to 1 - \frac{12}{\pi^2}\bigl(2x^2+2x+1\bigr)e^{-2x},
  \quad
  \tanh x \to 1 - 2e^{-2x},
\label{eq:G-asymptotics}
\end{equation}
At small $x$, $\mathcal G(x)=(4/\pi^2)x^3+O(x^5)$ whereas
$\tanh x=x+O(x^3)$. The cubic statement is pointwise in energy.
It gives cubic suppression of the integrated response when the whole
relevant bounded spectrum is uniformly small compared with $k_BT$.
It cannot be applied by replacing a dispersing spectrum with its minimum
gap, or by interchanging limits in an unbounded Dirac continuum.
The same-energy single-mass limit of the thermal integral is
$\mathcal G$, not $\tanh$.

\section{Band transport formula and normalization}
\label{app:band-transport}

With the Berry orientation fixed to match the FHS convention in Methods,
define the unhalved band sum
\begin{equation}
  \kappa_{xy}^{\rm Nambu}
  = \frac{1}{\hbar T}\sum_{n}\!\int\!\frac{\dd^2k}{(2\pi)^2}\,
    \Omega^z_n(\bm{k})\!\int_{E_{n,\bm{k}}}^{\infty}\!\!\dd\epsilon\;
    \epsilon^2 \left(-\frac{\partial f}{\partial\epsilon}\right).
\label{eq:kubo-BdG-band}
\end{equation}
The derivative is written as $-\partial f/\partial\epsilon$ so the
lower-band curvature contributes positively to the specified plateau.
The physical coefficient for the full Nambu Brillouin zone is
$\kappa_{xy}=\kappa_{xy}^{\rm Nambu}/2$.
This notation distinguishes the unhalved auxiliary sum from the physical response.
The energy-magnetization correction is already incorporated in the
transport expression~\cite{QinNiuShi2011,SumiyoshiFujimoto2013,Zeng2026}.
\section{Derivation of the holonomy-resummed kernel}
\label{app:kernel}

\emph{This appendix derives the parity-odd kernel
Eq.~\eqref{eq:K-full} from the vacuum-polarization tensor via
Matsubara summation; the result underpins the finite-temperature CS
comparison envelope of Appendix~\ref{subsec:kernel}.}

We derive Eq.~\eqref{eq:K-full} following the method of
Refs.~\cite{GhoshKlinkhamer2017,Ghosh2026,Ghosh2026BEC}.  Start from the
vacuum-polarization kernel on the cylinder
$\R_\tau \times \R_x \times S^1_L$:
\begin{align}
  \pi_{\text{odd}}^{\mu\nu}&(p_r) \nonumber \\
  &= \frac{1}{L}\sum_{n=-\infty}^{\infty}\int\frac{\dd^2 l}{(2\pi)^2}\,
  \frac{\tr[\gamma^\mu(\slashed{l} + m_v)\gamma^\nu
    (\slashed{l} + \slashed{p} + m_v)]_{\text{odd}}}
       {(l_n^2 + m_v^2)((l_n + p_r)^2 + m_v^2)},
\label{eq:app-pi}
\end{align}
where $l_n = (\vec{l},\,(2\pi n + \vth_v)/L)$,
$p_r = (\vec{p},\,2\pi r/L)$.  The parity-odd trace for
two-component fermions in $(2{+}1)$D is
$\tr[\gamma^\mu \gamma^\nu \gamma^\rho] = 2i\epsilon^{\mu\nu\rho}$ in a Hermitian Euclidean representation. The overall orientation and continuation factors are absorbed consistently in $\chi_v$ and Eq.~\eqref{eq:Pi-odd}.

After Feynman parameterization with parameter $u \in [0,1]$, shifting
$l_\mu \to l_\mu - up_\mu$, and performing the two-dimensional
momentum integral, we obtain
\begin{equation}
  \int\frac{\dd^2 l}{(2\pi)^2}\,
  \frac{1}{(l^2 + \Delta_{v,u}^2 + \omega_{n,u}^2)^2}
  = \frac{1}{4\pi(\Delta_{v,u}^2 + \omega_{n,u}^2)},
\label{eq:app-l-integral}
\end{equation}
where $\Delta_{v,u}^2 = m_v^2 + u(1-u)\tilde{p}^2$ and
$\omega_{n,u} = (2\pi n + \vth_v + 2\pi r u)/L$.

The Matsubara-type sum is evaluated using the standard identity
\begin{equation}
  \frac{1}{L}\sum_{n=-\infty}^{\infty}
  \frac{1}{\Delta^2 + \left(\frac{2\pi n + \theta}{L}\right)^2}
  = \frac{1}{2\Delta}\,
    \frac{\sinh(L\Delta)}{\cosh(L\Delta) - \cos\theta}.
\label{eq:app-Matsubara}
\end{equation}
This identity is proved by contour integration: define
$f(z) = [\Delta^2 + ((2\pi z + \theta)/L)^2]^{-1}$ and evaluate
$\frac{1}{L}\sum_{n} f(n) = \oint \dd z\,
\pi\cot(\pi z)\,f(z)/(2\pi i L)$ by closing in the upper and lower
half-planes, picking up the poles of $f(z)$ at
$z_{\pm} = (-\theta \pm iL\Delta)/(2\pi)$.  Equivalently, a partial-fraction expansion of the cotangent gives two geometric terms whose constant parts overlap. Their sum is
\begin{align}
  \frac{1}{L}\sum_{n} f(n)
  &= \frac{1}{2\Delta}\left[
    \frac{1}{1 - e^{-L\Delta + i\theta}}
    + \frac{1}{1 - e^{-L\Delta - i\theta}} - 1
  \right] \notag \\
  &= \frac{1}{2\Delta}\,
    \frac{\sinh(L\Delta)}{\cosh(L\Delta) - \cos\theta},
\label{eq:Matsubara-proof}
\end{align}
where the last step uses
$e^{2x} - 2e^x\cos\theta + 1 = 2e^x(\cosh x - \cos\theta)$ together
with $e^{x} - \cos\theta = \cosh x - \cos\theta + \sinh x$.  The subtracted unity follows from this algebraic decomposition. No pole at infinity is needed. It gives the required limit $1/(2\Delta)$ as $L\Delta\to\infty$.  Substituting
$\theta = \vth_v + 2\pi r u$ and $\Delta = \Delta_{v,u}$ immediately
yields Eq.~\eqref{eq:K-full}.

\section{Derivation of the exact thermal kernel}
\label{app:exact-kernel}

\emph{This appendix derives Eqs.~\eqref{eq:kxy-exact-kernel}
and~\eqref{eq:kernel-G} from the band Kubo formula, so that the exact
finite-temperature transport kernel is self-contained.  It uses no input from
Refs.~\cite{Ghosh2026,Ghosh2026BEC} beyond the identification of the
$T\to0$ limit, and may be read independently of
Appendix~\ref{app:kernel}.}

Start from the unhalved transport sum in Eq.~\eqref{eq:kubo-BdG-band},
including energy magnetization. The Berry orientation is fixed in
Methods, and the thermal weight is positive. For the
particle--hole-symmetric two-band form used in the
thermal curves,
$E_{\pm,\bm{k}} = \pm E_{\bm{k}}$ with $E_{\bm{k}} > 0$, and the Berry
curvatures of the two bands are opposite,
$\Omega_{+,\bm{k}} = -\Omega_{-,\bm{k}} \equiv -\Omega_{\bm{k}}$.
Writing $f(\epsilon) = (e^{\epsilon/k_BT}+1)^{-1}$ and
$w(\epsilon) = \epsilon^2\,(-\partial f/\partial\epsilon)$, the two
band contributions combine into
\begin{align}
  \hbar T\,\kappa_{xy}^{\rm Nambu}
  &= \int\!\frac{\dd^2k}{(2\pi)^2}\,\Omega_{\bm{k}}
    \left[\int_{-E_{\bm{k}}}^{\infty}\!\!\dd\epsilon\;w(\epsilon)
        - \int_{E_{\bm{k}}}^{\infty}\!\!\dd\epsilon\;w(\epsilon)\right]
    \notag\\
  &= 2\int\!\frac{\dd^2k}{(2\pi)^2}\,\Omega_{\bm{k}}
     \int_{0}^{E_{\bm{k}}}\!\!\dd\epsilon\;w(\epsilon),
\label{eq:app-two-band}
\end{align}
where the second line uses $w(-\epsilon) = w(\epsilon)$.  The two tails
above $\pm E_{\bm{k}}$ cancel and what survives is the thermal weight
lying \emph{inside} the gap window $|\epsilon| < E_{\bm{k}}$.  Using
\begin{equation}
  -\frac{\partial f}{\partial\epsilon}
  = \frac{1}{4k_BT}\,\mathrm{sech}^2\!\left(\frac{\epsilon}{2k_BT}\right),
\label{eq:app-sech}
\end{equation}
and substituting $u = \epsilon/2k_BT$, the total thermal weight is
normalized by
\begin{equation}
  \int_0^{\infty}\!\dd\epsilon\;
  \epsilon^2\left(-\frac{\partial f}{\partial\epsilon}\right)
  = 2(k_BT)^2\!\int_0^{\infty}\!\!\dd u\,u^2\,\mathrm{sech}^2u
  = \frac{\pi^2}{6}(k_BT)^2,
\label{eq:app-normalization}
\end{equation}
using $\int_0^\infty u^2\,\mathrm{sech}^2u\,\dd u = \pi^2/12$.  As
$T \to 0$ the upper limit $E_{\bm{k}}/2k_BT \to \infty$ and the whole
weight is collected, so with
$\CBdG = (2\pi)^{-1}\!\int\!\dd^2k\,\Omega_{\bm{k}}$ and
$\hbar = h/2\pi$, Eq.~\eqref{eq:app-two-band} returns
\begin{equation}
  \frac{\kappa_{xy}^{\rm Nambu}}{T}\bigg|_{T\to0}
  = \frac{2}{\hbar T^2}\cdot\frac{\pi^2(k_BT)^2}{6}\cdot
    \frac{\CBdG}{2\pi}
  = \frac{\pi^2k_B^2}{3h}\,\CBdG .
\label{eq:app-nambu-plateau}
\end{equation}
This is the \emph{doubled} value, because
Eq.~\eqref{eq:kubo-BdG-band} sums over both bands of a
particle-hole-doubled Hamiltonian in which every physical excitation is
counted twice.  Halving it, as required by the Nambu bookkeeping of
Appendix~\ref{app:conventions}, gives the physical plateau
$\kappa_{xy}/T = (\pi^2k_B^2/6h)\CBdG$ of
Eq.~\eqref{eq:kxy-majorana-quantum}.  Equation~\eqref{eq:app-nambu-plateau} is an internal bookkeeping check. Comparison with Ref.~\cite{Zeng2026} also requires matching the retained physical sectors and the definition of its Chern number.

At finite temperature part of the weight leaks outside the window
$|\epsilon| < E_{\bm{k}}$ and is lost, so each momentum contributes the
surviving fraction
\begin{equation}
  \mathcal{G}(x)
  = \frac{\int_0^{x}u^2\,\mathrm{sech}^2u\;\dd u}
         {\int_0^{\infty}u^2\,\mathrm{sech}^2u\;\dd u}
  = \frac{12}{\pi^2}\int_0^{x}\! u^2\,\mathrm{sech}^2 u\;\dd u,
  \qquad x = \frac{E_{\bm{k}}}{2k_BT},
\label{eq:app-G-def}
\end{equation}
which is Eq.~\eqref{eq:kernel-G} in integral form and gives
Eq.~\eqref{eq:kxy-exact-kernel} after weighting by
$\Omega_{\bm{k}}$.  The closed form follows from two integrations by
parts.  Writing $\mathrm{sech}^2u = \dd(\tanh u)/\dd u$,
\begin{equation}
  \int_0^{x}u^2\,\mathrm{sech}^2u\,\dd u
  = x^2\tanh x - 2\!\int_0^{x}\!u\tanh u\,\dd u,
\label{eq:app-ibp1}
\end{equation}
and with $\tanh u = 1 - 2(e^{2u}+1)^{-1}$,
\begin{equation}
  \int_0^{x}\!u\tanh u\,\dd u
  = \frac{x^2}{2} + x\ln\!\left(1+e^{-2x}\right)
    - \frac{1}{2}\Bigl[\mathrm{Li}_2\!\left(-e^{-2x}\right)
    - \mathrm{Li}_2(-1)\Bigr],
\label{eq:app-ibp2}
\end{equation}
where $\mathrm{Li}_2(-1) = -\pi^2/12$.  Combining
Eqs.~\eqref{eq:app-ibp1} and~\eqref{eq:app-ibp2} and multiplying by
$12/\pi^2$ gives
\begin{equation}
  \mathcal{G}(x)
  = 1 - \frac{12}{\pi^2}\Bigl[\,x^2\bigl(1-\tanh x\bigr)
     + 2x\ln\!\left(1+e^{-2x}\right)
     - \mathrm{Li}_2\!\left(-e^{-2x}\right)\Bigr],
\label{eq:app-G-closed}
\end{equation}
which is Eq.~\eqref{eq:kernel-G}.  Expanding
Eq.~\eqref{eq:app-G-closed} at large $x$ with
$1-\tanh x = 2e^{-2x}+\mathcal{O}(e^{-4x})$,
$\ln(1+e^{-2x}) = e^{-2x}+\mathcal{O}(e^{-4x})$, and
$\mathrm{Li}_2(-e^{-2x}) = -e^{-2x}+\mathcal{O}(e^{-4x})$ gives the
first asymptotic form in Eq.~\eqref{eq:G-asymptotics}; expanding the
integral form at small $x$ with $\mathrm{sech}^2u \to 1$ gives
$\mathcal{G}(x)\to(4/\pi^2)x^3$.  Both limits, and the value
$\mathcal{G}(\infty)=1$, were verified numerically against direct
quadrature of Eq.~\eqref{eq:app-G-def}.

If the curvature-weighted distribution of $E_{\bm{k}}$ is
concentrated at a single energy, Eq.~\eqref{eq:kxy-exact-kernel}
collapses to $\mathcal{G}$ evaluated at one argument.  It does not become
$\tanh$ of that argument.  Equation~\eqref{eq:kxy-T} is therefore an
independent single-mass parity-odd comparison envelope.  The fitted
rescaling described in Appendix~\ref{app:envelope} quantifies when that
envelope approximates the exact Kubo result over a restricted
temperature interval.

\section{Conventions and the Nambu prefactor}
\label{app:conventions}

A chiral Majorana channel has half the heat capacity and central charge
of a chiral Dirac channel. In the full Brillouin-zone Nambu description,
the band sum has an overall factor $1/2$. This produces
$q_M=4.732\times10^{-13}$~W/K$^2$ for one Majorana mode and
$2q_M=9.464\times10^{-13}$~W/K$^2$ for one Dirac mode.
The factor follows from the redundant Nambu degrees of freedom; it
must be applied once, with all physical spin and valley sectors included.

Ref.~\cite{Zeng2026} uses the Dirac scale
$\kappa_0^{\rm Z}=\pi k_B^2T/(6\hbar)=\pi^2k_B^2T/(3h)$.
In our convention a total physical Chern number $C$ gives
$\kappa_{xy}/\kappa_0^{\rm Z}=C/2$ at low temperature.
Our unhalved auxiliary band sum instead returns $C$. Stating these
normalizations is necessary, but does not establish agreement between
different microscopic models. A comparison must also match the number
of retained sectors and any reduced-block convention. A disagreement
in physical channel count cannot be removed by changing the unit alone.

For the illustrative choice $T_c=300$~mK and
$2\Delta_0/(k_BT_c)=20$, $\Delta_0=0.2585$~meV. These values set a
scale for Fig.~\ref{fig:sixpanel}f; the gap ratio is not a spectroscopic
measurement in Ref.~\cite{Han2025}. The chosen $\Tstar=5T_c$ is also
illustrative and affects the full continuation curve, not just its
termination point. The absolute gapped low-temperature values are
\begin{align}
  \CBdG = 1\;(p+ip)&:
    \quad \frac{\kappa_{xy}}{T}\big|_{T\to 0}
    = 4.732\times10^{-13}\;\frac{\mathrm{W}}{\mathrm{K}^2},
\label{eq:kxy-C1}\\
  \CBdG = 2\;(\text{total})&:
    \quad \frac{\kappa_{xy}}{T}\big|_{T\to 0}
    = 9.464\times10^{-13}\;\frac{\mathrm{W}}{\mathrm{K}^2},
\label{eq:kxy-C2}
\end{align}
The second line refers to the total $C=2$ benchmark. Neither value
fixes a unique microscopic pairing symmetry. Above $T_c$, a spectral
ansatz and a long-lived quasiparticle assumption remain insufficient
without the corresponding transport calculation.

\section{Accuracy of the tanh envelope}
\label{app:envelope}

Table~\ref{tab:envelope} compares the two functions numerically. The exact column
uses the energy-resolved kernel, whereas the comparison uses
$\tanh[\Delta/(2k_BT)]$. A fitted scale
$\Delta_{\rm eff}/\Delta=0.837$ closely matches the sampled Dirac-cone
curve over a restricted interval. In the notebook this scale is the
mean of pointwise matching scales, not a global minimax fit. It is not
an identity between response functions.

The continuum integral uses an upper cutoff $E=600$ in units with
$m=1$ and normalization to the infinite-cutoff plateau. This accounts
for a small residual offset even at low temperature. The lattice
comparison has $\xi=-2t(\cos k_x+\cos k_y)-\mu_b$ with $t=\mu_b=1$
and $\Delta_0=0.8$. Its chemical-potential convention and filling differ
from the anchor in Methods. Its curvature distribution produces a
substantially different temperature dependence. A fitted $\tanh$ scale
therefore cannot be used as an independent determination of the gap.

The small-argument cubic law for $\mathcal G$ is local in energy.
It is uniform for a bounded spectrum taken to high temperature, but
not for an unbounded Dirac cone or merely a vanishing minimum gap.
At $T/\Delta=0.2$, the lattice and envelope values already
differ by about four percentage points of the plateau. The acceptable
temperature window should be set by the desired accuracy and the
actual excitation spectrum.

\begin{table}[H]
  \caption{Thermal Hall conductance normalized to its $T\to0$ plateau.
    ``Exact'' is Eq.~\eqref{eq:kxy-exact-kernel}; ``$\tanh$'' is
    Eq.~\eqref{eq:kxy-T} evaluated at the minimum BdG gap $\Delta$;
    ``$\tanh_{\rm eff}$'' uses the rescaled scale
    $\Delta_{\rm eff} = 0.837\,\Delta$, with $0.837$ estimated once over
    $T/\Delta\in[0.2,1.3]$ and held fixed thereafter.  Lattice parameters $t=1$, $\mu_b=1$, $\Delta_0=0.8$, $N_k=401$, with $\xi=-2t(\cos k_x+\cos k_y)-\mu_b$.}
  \label{tab:envelope}
  \begin{ruledtabular}
  \begin{tabular}{cccccc}
    & \multicolumn{3}{c}{Dirac cone}
    & \multicolumn{2}{c}{lattice $p+ip$} \\
    \cline{2-4}\cline{5-6}
    $T/\Delta$ & exact & $\tanh$ & $\tanh_{\rm eff}$ & exact & $\tanh$ \\
    \hline
    0.10 & 0.998 & 1.000 & 1.000 & 0.999 & 1.000 \\
    0.15 & 0.992 & 0.998 & 0.993 & 0.988 & 0.998 \\
    0.20 & 0.970 & 0.987 & 0.970 & 0.947 & 0.987 \\
    0.30 & 0.884 & 0.931 & 0.884 & 0.792 & 0.931 \\
    0.50 & 0.685 & 0.762 & 0.684 & 0.461 & 0.762 \\
    0.70 & 0.536 & 0.613 & 0.536 & 0.259 & 0.613 \\
    1.00 & 0.396 & 0.462 & 0.396 & 0.116 & 0.462 \\
  \end{tabular}
  \end{ruledtabular}
\end{table}


\section{Protection of the low-temperature plateau}
\label{subsec:protection}

The quantized coefficient rests on the chiral central charge of the
gapped phase, not on finite-temperature exactness of a charge response.
For a system adiabatically connected to the class-D BdG state,
$c_-=\CBdG/2$ fixes the zero-temperature coefficient. Weak interactions
that preserve that phase do not continuously change its value.
More general interacting phases can require a many-body description
rather than a single-particle BdG Chern number.

For the massive continuum determinant at fixed holonomy,
\begin{equation}
  R(\lambda,\vth) = 1 + 2\sum_{\ell=1}^{\infty}
  e^{-\ell\lambda}\cos(\ell\,\vth),
\label{eq:R-expansion}
\end{equation}
There is no isolated inverse-circumference term in this expansion:
\begin{equation}
  c_1 = 0
\label{eq:c1-zero}
\end{equation}
This is the restricted finite-size statement inherited from
Refs.~\cite{Ghosh2026,Ghosh2026BEC}. It does not exclude finite-size
corrections from gapless edges, nonideal contacts, or other device modes.
The six-width Wilson-loop sequence is a supporting model comparison,
rather than a proof of the absence of all algebraic contributions.

\section{Coleman--Hill theorem and its limits}
\label{app:coleman-hill}

The Coleman--Hill theorem concerns a zero-temperature Abelian gauge
theory under its gauge-invariance and analyticity assumptions
\cite{ColemanHill1985}. In the conventional Lorentz-invariant setting,
write the parity-odd self-energy as
\begin{equation}
  iD_{\mu\nu}^{-1}(k) \ni i\epsilon_{\mu\nu\lambda}k^\lambda
  \Pi_2(k^2).
\label{eq:app-CH-propagator}
\end{equation}
At one loop,
\begin{equation}
  \Pi_2(0)\big|_{\text{1-loop}}
  = \frac{\mu}{e_0^2}
    + \frac{1}{4\pi}\sum_{\text{spinors}} q_a^2\,
      \frac{m_a}{|m_a|}.
\label{eq:app-CH-oneloop}
\end{equation}
Here $\mu$ in the first term is the bare topological mass of the gauge
theory, not the electronic chemical potential used elsewhere.
The Ward identities and analyticity imply that vertices with more than
two gauge legs contain sufficient powers of external momenta to remove
higher-loop corrections to $\Pi_2(0)$. This is a non-renormalization
result distinct from, although analogous in spirit to,
Adler--Bardeen~\cite{AdlerBardeen1969}.

A fermion gap alone does not supply the assumptions at finite temperature.
Thermally occupied states can make the response nonanalytic at small
frequency and momentum. Explicit nonzero two-loop corrections occur
in thermal QED$_3$~\cite{Brandt2000}. The theorem therefore does not
protect a general $\tanh[\Delta(T)/(2k_BT)]$ coefficient, even at fixed
gap. A superconducting electromagnetic response also involves broken
charge $U(1)$ symmetry and collective vertices. It cannot be identified
with an ordinary minimally coupled Dirac polarization without further
work. Neither the finite-temperature envelope nor the above-$T_c$
continuation in this manuscript is claimed to be one-loop exact under
interactions. This restriction is also consistent with the distinction
between spatial compactification at zero temperature and physical
finite temperature in Ref.~\cite{Ghosh2026}.

\section{Finite-size corrections and radiative scope}
\label{app:protection-detail}

For the fixed-mass continuum comparison, the exponential structure gives
\begin{equation}
  \frac{\kappa_{xy}^{\rm env}}{T}(L)
  = \frac{\kappa_{xy}^{\rm env}}{T}(\infty)
    \left[1 + \mathcal{O}\!\left(e^{-L/\xi}\right)\right],
  \quad
  \xi = \frac{\hbar v_F}{\Delta}.
\label{eq:finite-size}
\end{equation}
The superscript ``env'' restricts this expression to the determinant-based
comparison coefficient. For a general gapped local model, corrections
may have the form $\sum_j A_j(L)e^{-L/\xi_j}$ with polynomial prefactors.
The edge level spacing itself can scale as $1/L$ because the edge is
gapless. These are different observables.

The Wilson-loop cycle winding is fixed topologically even on a narrow
cylinder. Its integrality does not prove a particular local finite-width
law. The observed RMS sequence is compatible with exponential
suppression over the sampled widths, but the available data do not
statistically exclude a power law or an exponential with a polynomial
prefactor. Fits are reported as finite-width diagnostics.

At finite temperature, interaction corrections to transport can arise
through self-energies, vertices, and collective modes. They cannot in
general be absorbed into $\Delta(T)$ alone. The exactness of
Eqs.~\eqref{eq:kxy-exact-kernel} and~\eqref{eq:branch-weight} refers
to the clean quadratic two-band problem. Protection of the low-temperature
plateau follows from the gapped phase and its chiral central charge,
with the experimental qualifications stated in the main text.

\section{Read--Green topology and the sign convention}
\label{app:read-green}

For a continuum chiral gap with a specified ultraviolet completion,
the degree of the normalized vector
$\bm E=(\mathrm{Re}\Delta_{\bm k},-\mathrm{Im}\Delta_{\bm k},\xi_{\bm k})$
is~\cite{Read2000}
\begin{equation}
  \mathcal{M} = \int\frac{\dd^2 p}{8\pi}\,
  \epsilon_{ij}\,
  \hat{\bm{E}}_{\bm{p}} \cdot
  \left(\partial_i \hat{\bm{E}}_{\bm{p}}
  \times \partial_j \hat{\bm{E}}_{\bm{p}}\right).
\label{eq:RG-winding}
\end{equation}
For $\Delta_{\bm k}\propto k_x+ik_y$, $\xi_{\bm k}=k^2/(2m)-\mu$,
and $m>0$, this degree is $\mathcal M=+1$ at $\mu>0$ and zero at
$\mu<0$. The occupied-band FHS orientation used in this manuscript
has $\CBdG=-\mathcal M$. Its positive anchor is therefore the complex
conjugate gap. The overall orientation is conventional, but must be
used consistently when comparing a phase texture, a numerical Chern
number, and a laboratory heat-flow direction.

For a fixed static BdG Hamiltonian, changing a nonzero gap amplitude
without closing the spectrum leaves this degree unchanged. This does
not prove that a phase-incoherent pseudogap has the same invariant.
A normal self-energy with a gap-like spectral feature is insufficient
to construct the anomalous Green function or its transport vertices.
Momentum-space topology can organize transitions more generally
\cite{KlinkhamerVolovik2004}, but the gap closing at $\mu=0$ in this
two-dimensional model should not be identified with a pair of stable
three-dimensional Fermi points.

\section{Gravitational response and the physical thermal quantum}
\label{app:grav-cs}

In the trace convention of Ref.~\cite{Ryu2012}, the fully regularized
two-dimensional phase has the canonical gravitational Chern--Simons
coefficient
\begin{equation}
  I_{\text{CS}}^{\text{grav}}
  = \frac{1}{4\pi}\cdot\frac{c}{24}
    \int \dd^3 x\,\epsilon^{ijk}\,\tr\!\left(
    \omega_i \partial_j \omega_k
    + \tfrac{2}{3}\omega_i \omega_j \omega_k\right),
\label{eq:grav-cs-detail}
\end{equation}
Here $c=c_-$ is the total chiral central charge, including orientation.
A chiral Dirac channel has $c=1$ and a Majorana channel has $c=1/2$.
The extra factor $1/2$ for an individual massive surface contribution
in Ref.~\cite{Ryu2012} is not an additional factor for a fully
regularized two-dimensional Majorana phase.
With the zero-temperature normalization $K_{\rm env}(0)=2\CBdG$,
Luttinger's thermal response framework gives
\cite{Luttinger1964,Volovik2003}
\begin{equation}
  \frac{\kappa_{xy}}{T}\bigg|_{T\to0} = \frac{\pi^2 k_B^2}{12h}\,\Kem(0)
  = \frac{\pi^2 k_B^2}{6h}\,\CBdG,
\label{eq:kxy-from-grav}
\end{equation}
The equality is a statement about the low-temperature limit.
It does not imply a universal proportionality between electromagnetic
and gravitational responses at finite temperature. Their distinction
is why the finite-temperature transport calculation is required.

\section{Backgrounds and discriminating measurements}
\label{subsec:onset}
\label{app:backgrounds}

\paragraph{Impurity-induced response.}
Branch-conversion scattering can generate an anomalous bulk thermal
Hall response that is sensitive to the impurity potential
\cite{Ngampruetikorn2020,YilmazYip2020}. In the models considered there,
point scatterers distinguish unit winding from higher winding, whereas
finite-range scatterers can activate additional channels. This response
need not follow the linear Chern staircase. If a full quasiparticle gap
survives, thermally activated bulk contributions vanish at low
temperature. Disorder dependence and temperature dependence then help
separate them from the plateau. Strong pair breaking or subgap impurity
bands invalidate that simple freeze-out argument.

\paragraph{Bogoliubov Fermi surfaces.}
The eight-band calculation of Ref.~\cite{YangZhang2025} finds gapless
regions near the superconducting boundary and near its mean-field
transition temperature. Ref.~\cite{Zeng2026} shows how these regions
alter thermal transport and enhance Nernst responses. Our controlled
map illustrates the distinction between the surviving isolated-band
invariant and the loss of a filled lower-band bundle. Both works must
be compared using the same physical sectors and the same Nambu
normalization. A residual electronic $\kappa_{xx}/T$, spectroscopy,
and noninteger transverse response can support a gapless interpretation.
None of these signatures by itself uniquely identifies a Bogoliubov
Fermi surface. An apparently integer transverse signal alone also
cannot prove the absence of low-energy bulk states.

\paragraph{Fluctuations, normal carriers, and phonons.}
Chirality-polarized fluctuations can generate Hall and Nernst terms,
including clean-limit contributions proportional to $\tau^2$
\cite{SumiyoshiFujimoto2014}. This is a reason to resolve separate
mechanisms above $T_c$, rather than assume quantized continuation.
Measure temperature, weak-field, frequency, and disorder dependence
in the same prepared domain. A pairing-suppressed control estimates
the parent electronic background, with due allowance for changes
caused by the suppressing field or gate. Contact asymmetry is often
domain even, but phonon or substrate Hall signals coupled to magnetic
order can also be domain odd. Domain reversal is therefore a useful
control, not a complete subtraction of every background.

\section{Berry-ring comparison and its limitations}
\label{subsec:ring}

The isotropic two-band Berry curvature is largest at
\begin{equation}
  k_\Omega=\left(\frac{2D^2}{v_N^2}\frac{N-1}{N+2}\right)^{1/2N},
\label{eq:BRF-radius}
\end{equation}
Here $v_N$ is the coefficient of $k^N$ in the off-diagonal parent-band
Hamiltonian; its units are energy times length$^N$. With the corresponding
isotropic parametrization, $v_N=D/k_0^N$. The radius follows by
maximizing $k^{2N-2}/(D^2+v_N^2k^{2N})^{3/2}$.
It is a parent-band scale and is not itself a gap-closing condition.

A nondegenerate circular pocket over the cited tetralayer density range
has $k_F=0.243$--$0.270$~nm$^{-1}$, below the model ring radius
$k_\Omega=0.378$~nm$^{-1}$. A semiclassical conversion of the higher
oscillation tones gives
\begin{equation}
  k_{\rm orb}=\sqrt{4\pi n_{\rm SdH}}\simeq0.34\text{--}0.37~\text{nm}^{-1},
\label{eq:korb-multitone}
\end{equation}
These are area-equivalent radii. For an anisotropic orbit, equality of
area does not mean its boundary lies on a circular ring. Magnetic
breakdown can further invalidate their interpretation as individual
pocket radii~\cite{Kalantre2026,Zhao2026}. The numerical proximity is a
motivation to investigate nonuniform Berry curvature, not evidence that
a small gate change must produce a topological transition. No invariant
calculated here depends on this conversion.
\subsection{Berry-ring placement: parameters and uncertainties}
\label{app:ring}

\emph{This subsection records the densities, parameters, and sensitivity
of the comparison between the occupied momentum-space radii and the
Berry-curvature ring quoted above.}

A nondegenerate circular pocket has $k_F=\sqrt{4\pi n_e}$, which
reproduces the value $k_F\simeq0.25$~nm$^{-1}$ quoted for
$n_e=0.5\times10^{12}$~cm$^{-2}$ in Ref.~\cite{Kalantre2026}.  Over the
R4G superconducting window $n_e=0.47$--$0.58\times10^{12}$~cm$^{-2}$
measured there, $k_F=0.243$--$0.270$~nm$^{-1}$, against
$k_\Omega=0.378$~nm$^{-1}$ from Eq.~\eqref{eq:BRF-radius} at the
interlayer potential $u_D=42.8$~meV used in their modelling.  Across
the wider Han density range and the R5G parameters of
Ref.~\cite{Patri2025} the disk lies inside the ring by a factor of
$1.4$--$2.3$.  In the Hartree--Fock analysis of
Ref.~\cite{Kalantre2026}, the circular quarter metal is favoured over
the annular state because its Fermi surface lies inside the
Berry-curvature ring. There the Bloch overlaps stay near unity and the
exchange energy is gained. An annular sea instead has its inner and
outer spinors rotated relative to one another by the enclosed
curvature.  This supports the usefulness of a circular reference state in that model. It does not establish a circular Fermi surface throughout the measured superconducting dome.

The multitone tones used in Eq.~\eqref{eq:korb-multitone} are reported
at $n_{\rm SdH}\simeq0.9\times10^{12}$~cm$^{-2}$ with a partner
separated by $\approx0.2\times10^{12}$~cm$^{-2}$; the separation is
quoted from a supplementary figure of Ref.~\cite{Kalantre2026}.  A
$\pm0.1\times10^{12}$~cm$^{-2}$ uncertainty in the upper tone moves
$k_{\rm orb}$ across $k_\Omega$ in either direction. This does not
change the conclusion of Sec.~\ref{subsec:ring}, which is that the
orbits lie near the ring rather than on a particular side of it.  As noted there, the semiclassical reading of the tones is
itself under discussion~\cite{Zhao2026}, and none of the invariants
computed in this work depends on it.

\section{Crossover diagnostics and the pair lifetime}
\label{app:crossover}

Within a two-gap spectral ansatz, one may have
\begin{equation}
  \Delta(T)|_{T>T_c}=\Dpg(T)\neq0,
  \qquad T_c<T<\Tstar,
\label{eq:Delta-above-Tc}
\end{equation}
A frequently used normal self-energy has the BCS-like form
\begin{equation}
  \Sigma(k)\simeq-\bigl(\Dsc^2+\Dpg^2\bigr)G_0(-k)
  \equiv-\Delta^2G_0(-k),
\label{eq:selfenergy-BCS-form}
\end{equation}
Momentum dependence of the gap form factor is suppressed in this
notation; for chiral pairing the corresponding $|\varphi_{\bm k}|^2$
must be retained. The comparison envelope below $T_c$ is
\begin{equation}
  \Kem(T)=2\CBdG\tanh\!\left[
  \frac{\sqrt{\Dsc^2(T)+\Dpg^2(T)}}{2k_BT}\right].
\label{eq:Kem-full}
\end{equation}
Here $K_{\rm env}$ remains the defined envelope, not a calculated
interacting electromagnetic or gravitational response. Above $T_c$,
phase averaging removes the anomalous expectation value. Recovering
a topology-sensitive transport term requires a dynamical Green function
and consistent vertices. A finite pair lifetime broadens its poles,
particularly as the spectral gap decreases. No lifetime correction is
computed by the present notebooks.

For the nondegenerate reference density and the quoted orbital
Ginzburg--Landau length from Ref.~\cite{Han2025},
\begin{equation}
  k_F \xi_{\rm GL} = 3.4\text{--}8.2
  \quad \text{for} \quad
  n_e = 0.4\text{--}0.85\times 10^{12}\,\text{cm}^{-2},
\label{eq:kFxi}
\end{equation}
This small product motivates crossover questions~\cite{Chen2024RMP}.
It is not a measurement of pair size. The orbital $\xi_{\rm GL}$ need
not equal the microscopic pair correlation length, and the conversion
$k_F=\sqrt{4\pi n_e}$ assumes a single nondegenerate circular pocket.
The normal-state reconstruction and degeneracy must be checked before
using it as a quantitative crossover criterion.

The BKT transition separates phase coherence from pair formation
\cite{Kosterlitz2016}. It permits, but does not quantify, a regime with
pair correlations above $T_c$. General statements that arbitrarily weak
attraction binds in two dimensions concern an appropriate attractive
$s$-wave problem. They do not remove the threshold for an odd-angular-
momentum two-body channel. No estimate of $\Tstar/T_c$ follows from
that argument for the present chiral state. The illustrative choices
$\Tstar/T_c=1.45$ and $5$ in the two figures are consequently independent
plot parameters. Spectroscopy and fluctuation measurements are needed
to determine whether a pseudogap exists and what scale it has.

The six-panel illustration stitches a BCS-like gap below
$T_c$ to a separate nonzero pseudogap above it. Its discontinuity is a
property of that ansatz, not evidence for a physical transition or a
smooth transfer between the two gaps. The separate framework figure
uses continuous illustrative profiles. Neither curve is a prediction
for the full measured response. The decomposition in
Eq.~\eqref{eq:kxy-decomposition} retains this distinction.

\section{Gate sweeps and extensions to other devices}
\label{subsec:D-field}
\label{subsec:extensions}

A continuous deformation of the gapped BdG Hamiltonian can change its
Chern number only through a gap closing. In the vortex description,
a zero of the pairing field crosses $\xi_s=0$. The disk proxy gives
$1\to0$ and the annular pair gives $0\to2\to0$. Nonuniform curvature
can also select higher odd sectors in suitable microscopic models
\cite{Patri2025}. These examples motivate simultaneous measurements of
thermal Hall response, longitudinal heat transport, and the gap during
a gate sweep. The correspondence requires following one homogeneous
phase. A first-order switch or domain repopulation need not exhibit a
continuous quasiparticle closing along the observed branch.

Superfluid stiffness supplies an independent low-energy probe.
A fully gapped state has activated quasiparticle depletion. Nodes can
produce a power law, including a linear term in the clean model of
Ref.~\cite{Patri2025}. The exponent and transition location depend on
the gap structure; neither follows from $k_F=k_\Omega$ alone.
Kinetic-inductance measurements in twisted graphene
\cite{banerjee2025superfluid} and local Meissner imaging in
rhombohedral graphene~\cite{ZhangMeissner2026} demonstrate relevant
experimental routes.

Octalayer graphene provides a distinct comparison. Ref.~\cite{Okounkova2026}
reports adjacent superconducting and anomalous-metal regimes with
contrasting parent degeneracies. The superconducting pocket admits a
BKT scale near $61$~mK. Representative orbital length scales are about
$300$~nm and $30$~nm. Using the stated half-metal estimate
$k_F=0.29$~nm$^{-1}$ gives $k_F\xi\simeq87$ and $9$. The second
number reuses that reference $k_F$ even though the anomalous metal
has a different reported degeneracy. It is an order-of-magnitude
comparison, not a measured pair size or a common microscopic Fermi
momentum. The threshold discussion in Ref.~\cite{Chen2024RMP} should
be applied with these distinctions in view.

The absence of an anomalous Hall response in the half-metal parent
suggests a different symmetry setting from the orbital ferromagnet of
Refs.~\cite{Han2025,Dutta2026}. It does not, by itself, force the
superconducting Chern number to vanish. A fully gapped chiral state
would have the corresponding Majorana plateau; an anomalous metal has
no such requirement. Measuring both regimes on one device would test
these alternatives without assuming the answer from parent polarization.

The thermal relation also applies to other layer numbers whenever the
same class-D and transport conditions hold. Proposed hexalayer states
and $(1{+}n)$ structures offer possible extensions. For the latter,
the $\CBdG=2n-1$ examples of Ref.~\cite{LeNir2026} imply
$(2n-1)\times4.73\times10^{-13}$~W/K$^2$. A reproducible staircase
would constrain higher-Chern superconductivity, but its location must
be determined by the actual pairing texture and occupied bands.
\section{Warped, finite-momentum model and its numerical map}
\label{app:warped}

\emph{This appendix specifies the model with a $C_{3z}$-symmetric normal dispersion used in
Sec.~\ref{subsec:warped} and records the numbers quoted there.}

The annular model of Appendix~\ref{subsec:annulus} is regularized on a square lattice. To preserve exact threefold symmetry in the normal dispersion,  the warped calculation uses a triangular-lattice regularization
built from three unit vectors $\hat{b}_i$ at $120^\circ$.  With
$d_i=\bm{k}\cdot\hat{b}_i$,
\begin{align}
  \rho^2(\bm{k}) &= \tfrac{4}{3}\Bigl[3-\textstyle\sum_i\cos d_i\Bigr]
    \;\to\; |\bm{k}|^2 ,
    \nonumber\\
  g_3(\bm{k}) &= \textstyle\sum_i\sin d_i
    \;\to\; -\tfrac{1}{8}|\bm{k}|^3\cos3\theta ,
    \nonumber\\
  (s_x,s_y) &= \tfrac{2}{3}\textstyle\sum_i\hat{b}_i\sin d_i
    \;\to\; (k_x,k_y) ,
\label{eq:warped-defs}
\end{align}
so that $\rho^2$ is $C_3$-symmetric and even in $\bm{k}$ while $g_3$ is
$C_3$-symmetric and odd, which is precisely the structure of trigonal
warping.  The normal dispersion and gap are
\begin{align}
  \varepsilon(\bm{k}) &=
    \bigl(\rho^2-r_{\rm in}^2\bigr)\bigl(\rho^2-r_{\rm out}^2\bigr)
    + w\,g_3(\bm{k}),
    \nonumber\\
  \Delta(\bm{k}) &= \Delta_0\,(s_x+i\chi_\Delta s_y)
    (s_x-q+i\chi_\Delta s_y)(s_x+q+i\chi_\Delta s_y),
\label{eq:warped-model}
\end{align}
with $r_{\rm in}=0.20$, $r_{\rm out}=0.55$, $\Delta_0=2.0$ and
$\chi_\Delta=-1$; $w$ is the warping strength and $q$ is the gap-texture
parameter entering the periodic coordinate $s_x$.  Near the zone centre
$q$ coincides with the radial position of the primary vortex pair, but
that identification is not exact near the lattice boundary.
Setting $w=0$ and $\bm{Q}=0$ gives the same local three-zero construction as Eq.~\eqref{eq:gap-finite-vortex}, with a different lattice regularization and parameters. The imposed vortex pair itself selects an axis, so the full paired state need not preserve $C_3$.  Finite pair momentum enters through
Eq.~\eqref{eq:xi-sa}.  Periodicity of every model function under the
reciprocal lattice dual to $\{\hat{b}_i\}$ is verified numerically
before any Chern number is computed.

\paragraph{Warping does not enter $\xi_s$ at $\bm{Q}=0$.}
Since $g_3$ is odd, at $\bm{Q}=0$ it contributes only to $\xi_a$.
Numerically, raising $w$ from $0$ to $3$ changes $\xi_s$ by at most
$3.6\times10^{-16}$ on a $301\times301$ grid while $\max|\xi_a|$ grows
from $0$ to $7.79$.  The warping contribution to the pairing-relevant set $\{\xi_s<0\}$ enters at order $w|\bm{Q}|$. A fixed $\bm Q$ selects a direction, so the deformation is not generally threefold symmetric; at $w=3$ and $|\bm{Q}|=0.10$ that distortion reaches
$0.30$ in units of $\varepsilon$.

\paragraph{Numbers quoted in Sec.~\ref{subsec:warped}.}
The vortex-crossing scan uses $30$ values of $q$ from $0.06$ to
$0.66$, refined on either side of each crossing, at
$w\in\{0,1,2,3\}$. The Chern number uses a $301\times301$ grid and
the gap minimum a $601\times601$ grid.  At every warping strength
$\max|C_{\rm BdG}^{\rm FHS}-Q_{\rm analytic}| \le 6.7\times10^{-16}$
and $\max|F_{\rm plaq}| = 2.60 < \pi$.  The fraction of the scan
satisfying Eq.~\eqref{eq:gapped-condition} is $1.00$, $0.73$, $0.43$ and
$0.27$ at $w=0,1,2,3$.  The $(w,|\bm{Q}|)$ map covers
$w\in[0,6]$ and $|\bm{Q}|\in[0,0.30]$ on a $25\times21$ grid at
$q=0.35$; the Chern number is $2$ at all $525$ points and the gapped
fraction of the map is $0.45$.  At $|\bm{Q}|=0.1\,k_F$, the value
reported in Ref.~\cite{YangZhang2025} with $k_F$ taken as $r_{\rm out}$,
the finite-grid estimate is positive at $w=0,1,2,3$. Continuous minimization at $w=3$ gives $\delta_{\rm BdG}\simeq-1.30\times10^{-4}$, compared with the grid estimate $+1.10\times10^{-4}$. The points at $w=4$ and $6$ are also gapless. The reported gapped fractions retain the original grid classification; points near the boundary require refinement. The reference is a comparison of momentum scales, not a material calibration of $w$.


\section{Real-space transport calculations}
\label{app:kwant}

\emph{This appendix specifies the real-space calculations of
Sec.~\ref{subsec:disorder} and Figs.~\ref{fig:disorder}
and~\ref{fig:walls}, and records their complete data.}

\begin{figure}[t]
  \centering
  \includegraphics[width=0.86\textwidth]{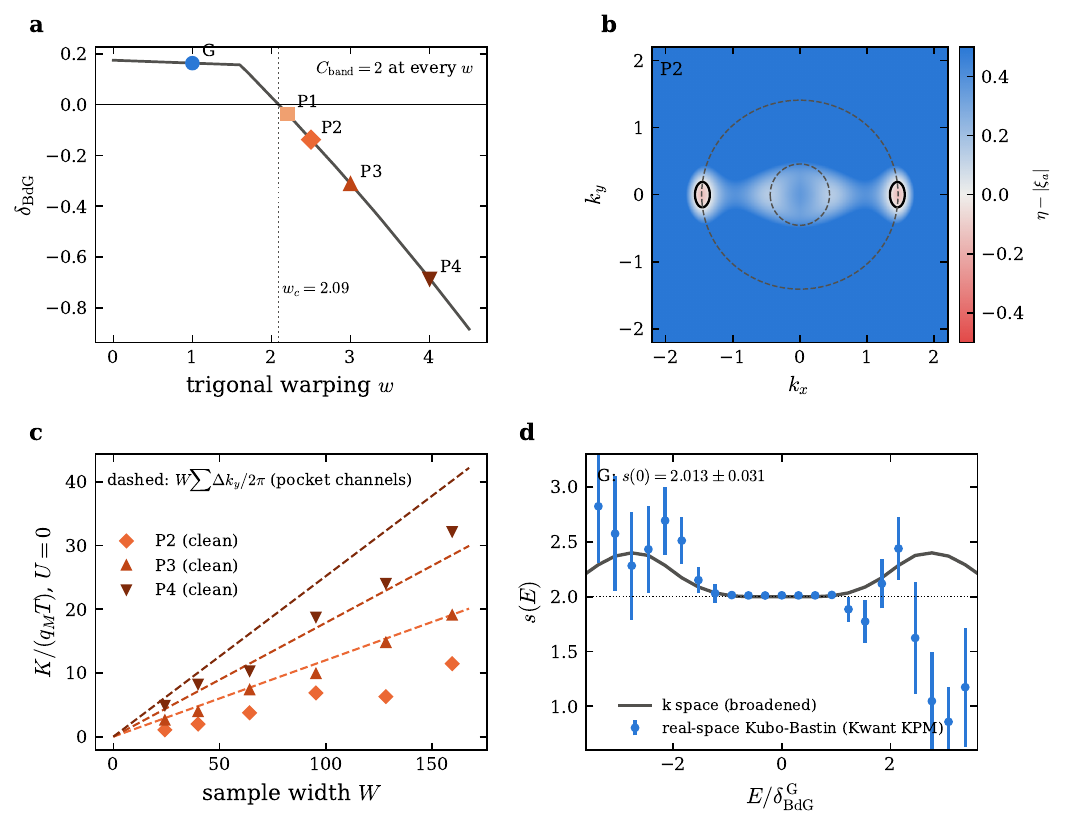}
  \caption{Lattice-resolved model and real-space bridge.
(a) Gap criterion of Eq.~\eqref{eq:gapped-condition} against warping at
$Q_x=0.1\,r_{\rm out}$; the band invariant is two throughout and
pockets open at $w_c=2.09$. (b) $\eta-|\xi_a|$ at P2; black contours
enclose the two pockets and dashed contours mark $\xi_s=0$. (c) Clean
heat conductance against sample width for P2--P4; dashed lines give the
pocket channel count $W\sum\Delta k_y/2\pi$. (d) Real-space Kubo--Bastin
Hall function of G from the kernel polynomial method, compared with the
k-space Hall function broadened to the same resolution; inside the gap
$s(0)=2.013\pm0.031$.}
  \label{fig:kwant-model}
\end{figure}

\paragraph{Lattice-resolved parameters.}
The parameters of Appendix~\ref{app:warped} give an occupied annulus
only $E_F\simeq0.017$ deep in a band about $34$ wide, with
$|\Delta|\gg E_F$. Disorder strong enough to produce a mean free path
shorter than accessible samples would then exceed $E_F$. The real-space
study keeps the functional form of Eq.~\eqref{eq:warped-model} and
rescales it in momentum by about $2.45$: $r_{\rm in}=0.45$,
$r_{\rm out}=1.35$, $q=0.90$, $\Delta_0=1.0$, $\chi_\Delta=-1$, and
$Q_x=0.1\,r_{\rm out}$. The ratio $q/r_{\rm out}$ is unchanged. This
set has $E_F=0.656$ and a minimum pairing scale $\eta_{\min}=0.265$ on
the Fermi sheets, so $\Delta/E_F<1$ as in the BCS regime. The Fermi
velocity is $v_F=2.63$, the coherence length $v_F/\eta_{\min}=9.9$, and
the normal density of states $\nu_N=0.100$ per site. It keeps $C_{\rm band}=2$
at every warping and is gapped up to $w_c=2.09$. Beyond the onset, two
pockets form on the $\pm\bm Q$ axis at $|\bm k|\simeq1.45$, as in
Fig.~\ref{fig:warped}a. Table~\ref{tab:kwant-points} lists the five
points. The same code reproduces the map of Appendix~\ref{app:warped},
with $C=2$ at all 525 points and gapped fraction $0.4533$.
Figure~\ref{fig:kwant-model}a,b shows the gap and a pocket map.

\begin{table}[H]
  \centering
  \caption{Points of the real-space study ($Q_x=0.1\,r_{\rm out}$).
    $k_p$ is the area-equivalent pocket radius and $\lambda_p=2\pi/k_p$.
    The last column is the clean $T\to0$ value of
    Eq.~\eqref{eq:curvature-blocking} in units of $q_M$.}
  \label{tab:kwant-points}
  \begin{ruledtabular}
  \begin{tabular}{cccccc}
    point & $w$ & $\delta_{\rm BdG}$ & $k_p$ & $\lambda_p$ & $\kappa_{xy}/(Tq_M)|_{T\to0}$ \\
    \hline
    G  & 1.0 & $+0.163$ & --    & --   & 2.000 \\
    P1 & 2.2 & $-0.037$ & 0.073 & 86   & 1.909 \\
    P2 & 2.5 & $-0.138$ & 0.144 & 44   & 1.712 \\
    P3 & 3.0 & $-0.312$ & 0.222 & 28   & 1.485 \\
    P4 & 4.0 & $-0.685$ & 0.336 & 19   & 1.222 \\
  \end{tabular}
  \end{ruledtabular}
\end{table}

\paragraph{Construction and validation.}
The hoppings follow from the exact inverse Fourier transform of
$\varepsilon(\bm k)$ and $\Delta(\bm k)$ on the triangular lattice,
with reconstruction error $1.4\times10^{-14}$ and odd pairing to
$10^{-16}$. In the gauge frame,
$c_{\bm r}\to e^{i\bm Q\cdot\bm r/2}c_{\bm r}$ makes the pairing
uniform and the Hamiltonian translation invariant; it is used on tori.
The laboratory frame carries the pair phase $e^{i\bm Q\cdot\bm R}$ at
each bond midpoint; it is used with leads and walls. The two frames
are unitarily equivalent to $1.9\times10^{-15}$, and torus spectra
equal the Bloch spectra on the commensurate momentum grid to
$1.2\times10^{-13}$. Table~\ref{tab:kwant-bench} lists the clean
benchmarks at G.

\begin{table}[H]
  \centering
  \caption{Clean benchmarks at G. Transmissions are Nambu traces in
    Majorana units. The cylinder results use samples of length 40.}
  \label{tab:kwant-bench}
  \begin{ruledtabular}
  \begin{tabular}{lcc}
    benchmark & value & expected \\
    \hline
    strip transmission (edges) & 2.0011 & $|C|=2$ \\
    cylinder transmission (bulk) & 0.0021 & 0 \\
    cylinder with two walls & 3.9986 & $2|C|=4$ \\
    scattering Chern number, $\chi_\Delta=\mp1$ & $\pm2$ & $\pm2$ \\
    $\min_\phi|\det r|$ & 0.9995 & 1 \\
    Bott index (torus of 16) & 2.0000 & 2 \\
    Kubo--Bastin $s(0)$ (flake of 60) & $2.013\pm0.031$ & 2 \\
  \end{tabular}
  \end{ruledtabular}
\end{table}

\paragraph{Observables.}
With $\mathcal T(E)$ the Nambu-trace transmission between the two
leads,
\begin{equation}
  \frac{K}{q_MT}=\frac{6}{\pi^2(k_BT)^2}\int_0^\infty\dd E\,E^2
  \left(-\frac{\partial f}{\partial E}\right)\mathcal T(E),
\label{eq:kwant-K}
\end{equation}
so that $g=\mathcal T(E\to0)$ at low temperature. For
$\mathcal T=1$ the right side equals one, the contribution of a single
chiral Majorana mode. The scattering Chern number is
\begin{equation}
  Q=\frac{1}{2\pi}\oint\dd\phi\;\partial_\phi\arg\det r(\phi),
\label{eq:kwant-Q}
\end{equation}
where $r$ is the reflection matrix of one lead and the boundary
condition $\psi\to e^{i\phi}\psi$ acts on the full Nambu spinor at the
seam of the cylinder~\cite{Fulga2012}. This treats $H_{\rm BdG}$ as a
class-A matrix, whose Chern number is $\CBdG$. Leads of decoupled
chains keep the mode content independent of $\phi$; with $24$ flux
points, $Q$ equals the FHS Chern number of Methods. The Bott
index~\cite{LoringHastings2010} uses all negative-energy states on
tori of $24$, $36$, and $48$ lattice units. The density of states uses
$3000$ Chebyshev moments on a torus of $96$ lattice units, a
resolution of $0.028$, below $\delta_{\rm BdG}/3$ at G. The real-space
Hall function uses $2000$ moments on a flake of $60$ lattice units,
sampled in its central half, and is calibrated on the square-lattice
anchor of Methods, which returns $1.006$. Figure~\ref{fig:kwant-model}d
compares it with the k-space Hall function.

Disorder is a uniform on-site potential in $[-U,U]$ in the $\tau_z$
channel, which preserves class D. The Born mean free path
$v_F/(2\pi\nu_NU^2/3)$ is $50$, $12.5$, and $3.1$ lattice units at
$U=0.5$, $1$, and $2$. The heat conductance uses square cylinders of
$L=24$, $40$, $64$, $96$, $128$, and $160$ lattice units with
$24$ disorder realizations each. The winding uses $L=64$ and $128$
with $12$ realizations, and the Bott index $8$ realizations per torus.

\paragraph{Gapped point.}
Table~\ref{tab:kwant-G} summarizes the quantization at G. The decay
length is obtained from $\langle\ln g\rangle$ over $L=40$--$96$.

\begin{table}[H]
  \centering
  \caption{Quantization at the gapped point G. $Q$ and
    $\min_\phi|\det r|$ refer to $L=128$; the Bott index is the mean
    over eight tori of $48$; $\rho(0)$ is normalized by $2\nu_N$.}
  \label{tab:kwant-G}
  \begin{ruledtabular}
  \begin{tabular}{cccccc}
    $U$ & decay length & $Q=2$ & $\min_\phi|\det r|$ & Bott & $\rho(0)/2\nu_N$ \\
    \hline
    0    & 4.01 & 1/1   & 1.000 & 2.00 & 0.000 \\
    0.25 & 4.01 & 12/12 & 1.000 & 2.00 & 0.000 \\
    0.5  & 4.02 & 12/12 & 1.000 & 2.00 & 0.000 \\
    1.0  & 4.06 & 12/12 & 1.000 & 2.00 & 0.000 \\
    1.5  & 4.22 & 12/12 & 1.000 & 2.00 & 0.001 \\
    2.0  & 5.37 & 12/12 & 1.000 & 2.00 & 0.028 \\
  \end{tabular}
  \end{ruledtabular}
\end{table}

\begin{figure}[t]
  \centering
  \includegraphics[width=0.9\textwidth]{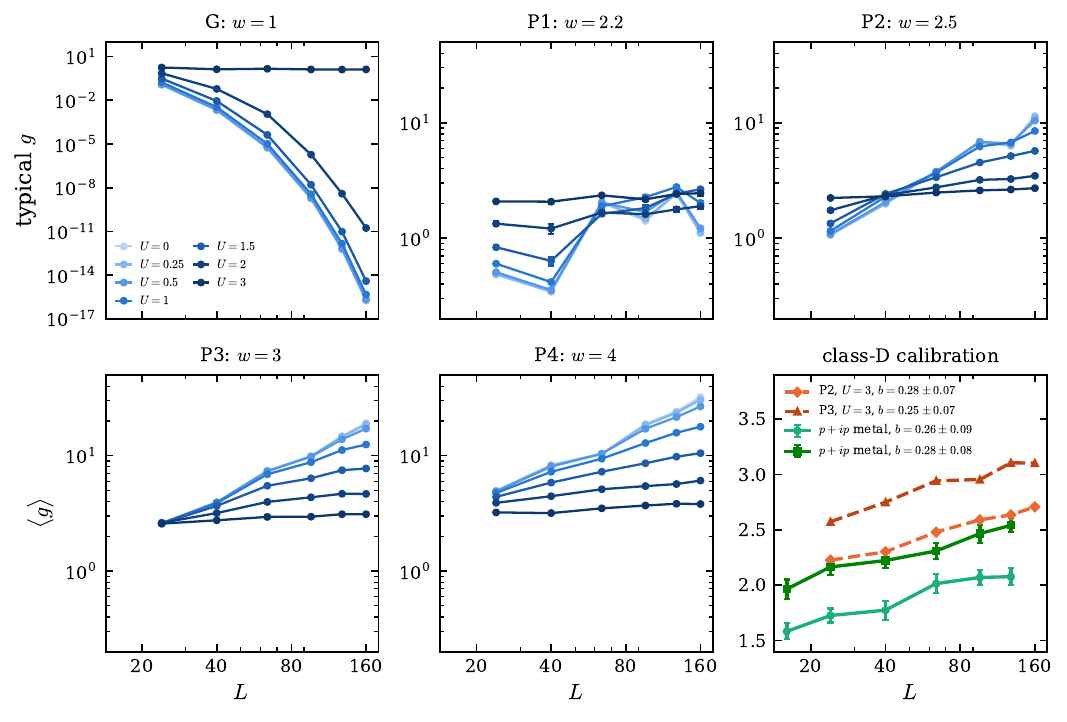}
  \caption{Complete disorder scaling. First five panels: typical
conductance at G and mean conductance at P1--P4 for all disorder
strengths, from light ($U=0$) to dark ($U=3$), with 24 realizations
per point. Last panel: calibration on the class-D thermal metal of the
square-lattice anchor with random chemical potential, compared with the
pocketed points at $U=3$; $b$ is the slope of $\langle g\rangle$
against $\ln L$.}
  \label{fig:kwant-scaling}
\end{figure}

\paragraph{Pocketed points.}
Table~\ref{tab:kwant-P} lists the conductance, the reflection
determinant, and the density of states. Without disorder the pocketed
bulk conducts ballistically, and $g$ grows linearly with the sample
width $W$. The slope is $58$, $68$, and $79$ percent of the pocket
channel count $W\sum\Delta k_y/2\pi$ at P2, P3, and P4, where
$\Delta k_y$ is the transverse extent of each pocket
(Fig.~\ref{fig:kwant-model}c).
Disorder reduces the conductance towards a diffusive value without
localizing it. For calibration we use the square-lattice $p+ip$ anchor
with a random chemical potential in $[\bar\mu-U,\bar\mu+U]$. Around
$\bar\mu=4t$ random domains of $C=\pm1$ form and their walls percolate,
the established route to the class-D thermal
metal~\cite{Chalker2001,Zakharov2025,SenthilFisher2000,EversMirlin2008}.
For $U=3t$ and pairing $t$ or $t/2$, $\langle g\rangle=a+b\ln L$ with
$b=0.26\pm0.09$ and $0.28\pm0.08$, while the gapped control at
$\bar\mu=2t$, $U=t$ gives $g\to0$. At $U=3$ the pocketed points give
$b=0.23\pm0.09$, $0.28\pm0.07$, $0.25\pm0.08$, and $0.48\pm0.06$ for
P1--P4, with bootstrap errors over $L\geq40$
(Fig.~\ref{fig:kwant-scaling}).

\begin{table}[H]
  \centering
  \caption{Pocketed points. $\langle g\rangle$ at $L=160$; median
    $\min_\phi|\det r|$ at $L=128$; $\rho(0)$ normalized by $2\nu_N$.}
  \label{tab:kwant-P}
  \begin{ruledtabular}
  \begin{tabular}{ccccc}
    point & $U$ & $\langle g\rangle$ & median $\min_\phi|\det r|$ & $\rho(0)/2\nu_N$ \\
    \hline
    P1 & 0   & 1.10  & $1.1\times10^{-2}$  & 0.099 \\
       & 0.5 & 1.22  & $2.3\times10^{-2}$  & 0.100 \\
       & 1   & 2.02  & $8.5\times10^{-3}$  & 0.110 \\
       & 2   & 1.89  & $8.7\times10^{-2}$  & 0.225 \\
    P2 & 0   & 11.46 & $1.4\times10^{-2}$  & 0.098 \\
       & 0.5 & 10.43 & $9.4\times10^{-4}$  & 0.122 \\
       & 1   & 8.49  & $4.9\times10^{-5}$  & 0.145 \\
       & 2   & 3.47  & $5.4\times10^{-3}$  & 0.266 \\
    P3 & 0   & 19.09 & $1.7\times10^{-4}$  & 0.125 \\
       & 0.5 & 17.14 & $1.5\times10^{-7}$  & 0.137 \\
       & 1   & 12.48 & $1.8\times10^{-7}$  & 0.153 \\
       & 2   & 4.65  & $5.3\times10^{-4}$  & 0.281 \\
    P4 & 0   & 32.22 & $1.7\times10^{-11}$ & 0.208 \\
       & 0.5 & 26.69 & $4.1\times10^{-12}$ & 0.182 \\
       & 1   & 17.79 & $1.2\times10^{-9}$  & 0.202 \\
       & 2   & 6.07  & $4.8\times10^{-4}$  & 0.357 \\
  \end{tabular}
  \end{ruledtabular}
\end{table}

\paragraph{Walls and textures.}
Bonds inside the time-reversed domain carry the complex-conjugate
hopping and pairing; bonds crossing a wall carry their real part. The
two walls lie at one and three quarters of the cylinder circumference,
here $64$ lattice units, so one wall carries heat in each direction.
Suppressed walls reduce the pairing to $0.3$ over a width of $3$;
metallic cores set it to zero over a width of $2$ or $4$.
Table~\ref{tab:kwant-walls} gives the wall conductance. Its excess over
$2|C|$ decays exponentially with wall length, over $8$--$15$ lattice
units for suppressed walls and $13$--$34$ for metallic cores. The
quantized value is limited by contact coupling, not by the wall. Metal
leads give $3.9999$ for the sharp wall, whereas leads of decoupled
chains with hopping $0.5$, $1$, and $2$ give $3.826$, $3.981$, and
$3.9987$. Random textures are Gaussian random fields with correlation
length $\ell_d$, thresholded at the reversed fraction. At $L=128$,
islands covering $25\%$ give $\langle g\rangle=0.20$ ($\ell_d=3$) and
$0.025$ ($\ell_d=6$), with $Q=2$ in all 16 textures at $L=64$ and median
$\min_\phi|\det r|$ of $0.94$ and $0.98$. Percolating textures covering
$50\%$ give $\langle g\rangle=2.71$ and $1.85$, with $Q=2$ in only 7 of
16 textures and median determinants of $0.005$ and $0.03$.

\begin{table}[H]
  \centering
  \caption{Heat conductance of a cylinder with two walls,
    $K_{\rm DW}/(q_MT)$, against wall length $L_x$. Clean values, with
    the values at $U=1$ in parentheses.}
  \label{tab:kwant-walls}
  \begin{ruledtabular}
  \begin{tabular}{lcccc}
    wall & $L_x=32$ & $64$ & $128$ & $192$ \\
    \hline
    erased      & 0.031 (0.053) & 0.000 (0.000) & 0.000 (0.000) & 0.000 (0.000) \\
    sharp       & 4.025 (4.084) & 4.000 (3.999) & 4.000 (3.999) & 4.000 (3.998) \\
    suppressed  & 4.271 (4.430) & 4.004 (4.051) & 3.999 (3.997) & 4.000 (3.997) \\
    metallic, 2 & 4.795 (4.442) & 4.338 (4.041) & 4.024 (4.003) & 4.002 (3.997) \\
    metallic, 4 & 6.150 (4.756) & 4.688 (4.068) & 4.017 (3.996) & 4.000 (3.991) \\
  \end{tabular}
  \end{ruledtabular}
\end{table}

\paragraph{Scales and scope.}
An illustrative mapping matches $\eta_{\min}$ to a BCS gap
$1.764\,k_BT_c$ with $T_c=0.3$~K and the coherence length to
$15.3$~nm. The energy unit is then $0.17$~meV and the lattice unit
$1.5$~nm, so $L=160$ corresponds to $0.25~\mu$m, the clean gap of G to
$28~\mu$eV, and the wall filtering lengths to $12$--$52$~nm. The pairing
field is imposed rather than self-consistent. Each domain contains one
valley-polarized orbital, the other valley entering by complex
conjugation, and scattering is elastic. These choices isolate the
mechanisms examined here; material-specific thresholds require the
multiband embedding discussed in the main text.

\section{Even-Chern scan: grid convergence and gap crossings}
\label{app:even-chern-numerics}

The primary zeros relevant to the occupied annulus lie on $k_y=0$.
The odd meshes used here do not sample that line. Other periodic image
zeros also exist, including on zone-boundary lines, but lie outside
the occupied region. A fine $k_y=0$ scan gives the plotted gap minima;
this is a line minimum, not an independently certified global minimum
at every parameter. The analytic conditions $\xi=\Delta=0$ establish
the actual closings at $r_{\rm in}$ and $r_{\rm out}$.

Near a crossing, a mesh can assign the wrong integer even though every
principal plaquette argument is strictly below $\pi$. The
$N_k=81$ example does so; representative assignments stabilize for
$N_k\geq121$. Near-$\pi$ flux is a warning, while mesh convergence and
the analytic crossing conditions decide the continuum sector.

\paragraph{Resolution study.}
An integer error of order $10^{-14}$ measures agreement between two
integer labels \emph{after} the mesh has assigned a topological sector.
It does not measure the distance from a sector error caused by
discretization. Stability under mesh refinement tests the sector assignment.
The margin $\pi-\max|F_{\rm plaq}|$ is a resolution diagnostic, not
an independent certificate. Table~\ref{tab:resolution-app} reports
the integer, line-gap estimate, and flux margins for eight radii.
Four radii bracket the two crossings. The assigned integer agrees at
all five tested meshes, $N_k=121,161,201,301,401$; four of these meshes
are displayed in the table.  Away from the crossings the margin is
close to $\pi$, between $3.03$ and $3.13$ at the finest mesh. At the
radii bracketing a crossing it is much smaller, as it must be. The
continuum Berry curvature is singular at a topological transition, and
the continuum invariant is undefined exactly at the gap-closing point.

The quantity that actually converges is the location of each crossing.
We locate the integer jump on a $41$-point radius grid. The inner
crossing lies at $0.4487$ for every mesh from $N_k=121$ upward. The
outer crossing drifts from $1.3413$ at $N_k=121$ to $1.3487$ at
$N_k=301$ and $401$. The exact values are $r_{\rm in}=0.4500$ and
$r_{\rm out}=1.3500$.  At the finest meshes both are recovered within the radius-grid spacing of $0.0025$. The coarsest outer-boundary estimate is less accurate.  Radii at which the mesh has not resolved the
transition are reported as such rather than assigned a phase.

\begin{table}[H]
  \centering
  \caption{Resolution study of the isotropic vortex-crossing scan
    ($r_{\rm in}=0.45$, $r_{\rm out}=1.35$, $\Delta_0=0.50$,
    $\chi_\Delta=-1$).  For each radius the assigned integer is
    identical at all five tested meshes; the four displayed mesh columns
    give the plaquette-flux margin $\pi-\max|F_{\rm plaq}|$.  $E_{\min}$ is the minimum quasiparticle
    gap from a dedicated fine scan along $k_y=0$, which contains the relevant primary zeros and is omitted by the odd two-dimensional meshes used here.
    Starred radii bracket a Fermi-sheet crossing.}
  \label{tab:resolution-app}
  \begin{ruledtabular}
  \begin{tabular}{ccccccc}
    radius & $\CBdG$ & $E_{\min}$
      & \multicolumn{4}{c}{$\pi-\max|F_{\rm plaq}|$} \\
    \cline{4-7}
     & & & $N_k{=}121$ & $161$ & $201$ & $401$ \\
    \hline
    0.1000            & 0 & $3.9\times10^{-2}$ & 2.905 & 2.990 & 3.035 & 3.112 \\
    0.3000            & 0 & $2.3\times10^{-2}$ & 2.733 & 2.911 & 2.958 & 3.069 \\
    0.4462\rlap{$^*$} & 0 & $6.7\times10^{-4}$ & 0.116 & 0.232 & 0.983 & 0.805 \\
    0.4600\rlap{$^*$} & 2 & $1.8\times10^{-3}$ & 0.850 & 1.079 & 1.862 & 1.791 \\
    0.8700            & 2 & $9.2\times10^{-2}$ & 3.029 & 3.064 & 3.094 & 3.129 \\
    1.3400\rlap{$^*$} & 2 & $1.3\times10^{-3}$ & 0.045 & 0.051 & 0.158 & 0.484 \\
    1.3600\rlap{$^*$} & 0 & $1.1\times10^{-3}$ & 0.479 & 0.207 & 0.285 & 0.495 \\
    1.4000            & 0 & $3.7\times10^{-3}$ & 0.928 & 0.484 & 0.746 & 1.336 \\
  \end{tabular}
  \end{ruledtabular}
\end{table}

A chirality-reversal check with $\chi_\Delta = +1$ gives $\CBdG = -2$ at
identical precision, confirming that $\CBdG = \pm 2$ is a physically
oriented invariant.

A third point concerns the scan range.  On the periodic lattice the
zeros of Eq.~\eqref{eq:gap-finite-vortex} occur wherever
$\sin k_x = \pm\sin q$, so each finite-momentum vortex has an image at
$k_x = \pi - q$.  For $2\sin(q/2) \lesssim 1.40$ the images lie far
outside the annulus and play no role, but they approach the primary
zeros as $q \to \pi/2$ and merge with them at the upper end of the
scan.  The scan is therefore terminated at $2\sin(q/2) = 1.40$, and no
statement is made about larger pair radii, where the periodic
regularization no longer represents a single well-separated pair.
The gap-phase texture is shown in Fig.~\ref{fig:gap-phase}.
Appendix~\ref{app:supporting} collects the square-lattice anchor and
the four annular benchmark cases in Tables~\ref{tab:pip-chern-app}
and~\ref{tab:four-cases-app}, respectively.

\section{Single-particle validation: methods and per-parameter data}
\label{app:single-particle}

We first verify the topological prefactor $\CBdG$ at the single-particle
level using the gauge-invariant Fukui--Hatsugai--Suzuki~\cite{Fukui2005}
discretisation.  The primary lattice representative is the spinless
$p+ip$ superconductor on a square lattice, Eq.~\eqref{eq:pwave-bdg},
with $t = 1$, $\mu = 1$, and $\Delta_0 \in \{0.2, 0.3, \ldots, 0.8\}$.
For $0 < \mu < 4t$ the system is in the weak-pairing topological
phase~\cite{Read2000}.  The FHS calculation on an $81\times 81$
$k$-grid gives $\CBdG = 1$ at every scan point, with the integer error
bounded by $10^{-14}$ in every run. The minimum direct band separation
$2E_{\min}$ grows monotonically from $0.347$ at $\Delta_0 = 0.2$ to
$1.346$ at $\Delta_0 = 0.8$, which rules out an accidental gap
closing.  Full
per-$\Delta_0$ data are tabulated in Appendix~\ref{app:supporting}
(Table~\ref{tab:pip-chern-app}).

To model the displacement-field-tuned Lifshitz transition of the
quarter-metal Fermi surface, we construct a lattice-regularized proxy
that captures the essential topological ingredients without a
self-consistent solution of the continuum model,
\begin{equation}
  \xi_D(\bm{k}) =
  \bigl[s(\bm{k}) - s_0(D)\bigr]^2 - \bar\mu,
  \quad
  s(\bm{k}) = 2(2 - \cos k_x - \cos k_y),
\label{eq:xi-proxy}
\end{equation}
with $s_0(D) = \sqrt{\bar\mu}\,D/D_c$, $\bar\mu = 0.36$, and
$D_c = 40$~meV as the proxy Lifshitz scale.  For $D < D_c$ the
occupied region is a disk centred on $\bm{k} = 0$. The central $p+ip$
vortex then lies inside the Fermi sea, so $\CBdG = 1$. For $D > D_c$
the region becomes an annulus and the central vortex is excluded, so
$\CBdG = 0$ in the minimal model.  The FHS scan over 15
displacement-field values in $D \in [10, 80]$~meV confirms this
picture (Fig.~\ref{fig:sixpanel}b,c). $\CBdG$ jumps $1 \to 0$ at
$D = D_c$, while the minimum quasiparticle gap falls by an order of
magnitude: from $0.55$ far from the transition to $0.057$ at
$D = 39$~meV and $0.048$ at $D = 41$~meV, against a pairing scale
$\Delta = 0.35$.  The scan brackets rather than resolves the closing,
because no grid point lies exactly at $D_c$. This softening is what a
finite scan of a topological transition driven by a gap closing at
$\bm{k} = 0$ should show.  This illustrates the relation between topology and a Lifshitz reconstruction in the proxy. It does not locate a transition in a specific graphene device.

\section{Wilson-loop flux threading on finite circumferences}
\label{app:wilson-loop}

The notebook samples transverse momenta
$k_{y,m}=(2\pi m+\Phi)/L_y$. For each subband it evaluates a Bloch
Wilson loop along the periodic $k_x$ direction, unwraps its phase as
$\Phi$ varies, and sums the phases. The polarization is
$P(\Phi)=-(2\pi)^{-1}\mathrm{Im}\ln\det W(\Phi)$ with a continuous
phase branch. This is a band-geometric cylinder construction. It does
not explicitly diagonalize an open-$x$ ribbon or simulate an
interacting physical electromagnetic flux pump.

For $t=\mu=1$ and $\Delta_0=0.4$, the cycle winding is
$\Delta P=1$ for $L_y=4,6,8,10,12,14$. The RMS deviations from a
straight line are $0.0448$, $0.0277$, $0.0064$, $0.0149$, $0.0068$,
and $0.0023$. The nonmonotonicity is clear. A log-linear fit to the
last four raw points gives a scale $5.12$; the reported selected-envelope
fit gives $3.11$. The notebook does not fit an explicit oscillatory
functional form or perform a statistical power-law comparison.

For context, the continuum estimate is $v_F/\Delta_0=5$ in the stated
units. The limited width range supports a qualitative comparison,
not a precision extraction of the asymptotic correlation length.
Terms $e^{-\ell L/\xi}\cos(\ell\vartheta)$ do not oscillate with $L$
at fixed $\vartheta$. Oscillations require additional size dependence,
for example from lattice momenta or complex decay scales. The exact
Poisson expansion alone does not explain the observed width oscillations.
Table~\ref{tab:wilson-app} gives the full sequence.

\section{Tensor-network benchmark of the quadratic BdG ground state}
\label{app:dmrg}

The FHS and Wilson-loop calculations establish the single-particle
topology and finite-width structure.  The DMRG campaign serves a
different purpose: it verifies that the finite-cylinder implementation
faithfully represents the finite-cylinder ground state of the explicitly
paired quadratic Hamiltonian and reproduces its local and edge
observables.  Because the complex $p+ip$ pairing term is present in the
Hamiltonian, the calculation does not test spontaneous selection of
chirality.

The simulations use the spinless lattice $p+ip$ model of
Eq.~\eqref{eq:pwave-bdg} in TeNPy~\cite{Hauschild2024} (v1.1.0,
Python 3.11) with $\mathbb{Z}_2$ fermion-parity conservation. The bond
dimension is ramped as $\chi=64\to128\to256\to512$ over 25 sweeps,
with $t=1$, $\mu=1$, $V_{nn}=0$, and gap chirality $\chi_\Delta=+1$.  The
scan covers $L_x=8$,
$L_y\in\{4,6,8,10\}$ and $\Dpg\in\{0.2,0.3,\ldots,0.8\}$.  The notebook records archive paths for MPS states, observables, and sweep histories. The accompanying package contains the notebooks and summary CSV. Its MPS archive paths refer to files outside that package.

\begin{figure}[t]
  \centering
  \includegraphics[width=0.95\textwidth]{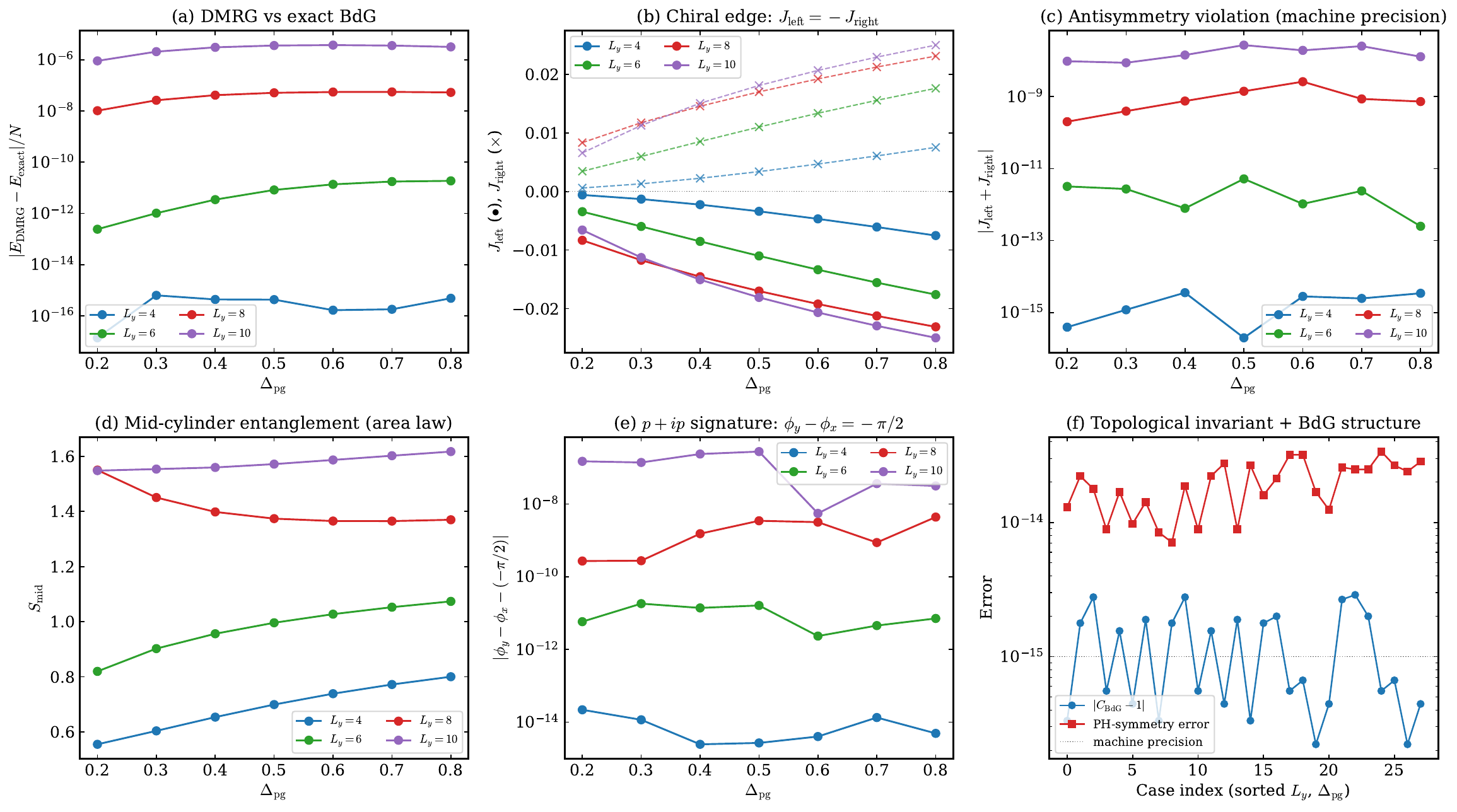}
  \caption{DMRG benchmark of 28 explicitly paired quadratic cylinders:
$L_x=8$, $L_y=4,6,8,10$, $\Dpg=0.2,\ldots,0.8$, $t=\mu=1$,
$V_{nn}=0$, and maximum bond dimension 512. Here $\Dpg$ is the
notebook's label for an imposed pair amplitude, not evidence for a
phase-incoherent pseudogap. (a) Energy error relative to exact
quadratic diagonalization. (b,c) Opposite edge currents and their
imbalance, with maximum relative imbalance about $1.5\times10^{-6}$.
(d) Entropy over four widths. (e) Recovery of the imposed pairing
phase, with maximum error about $2.8\times10^{-7}$~rad.
(f) Chern integrality and particle--hole residuals of the matching
Bloch and BdG matrices. These are not invariants extracted from the MPS.}
  \label{fig:dmrg-summary}
\end{figure}

\begin{figure}[t]
  \centering
  \includegraphics[width=0.9\textwidth]{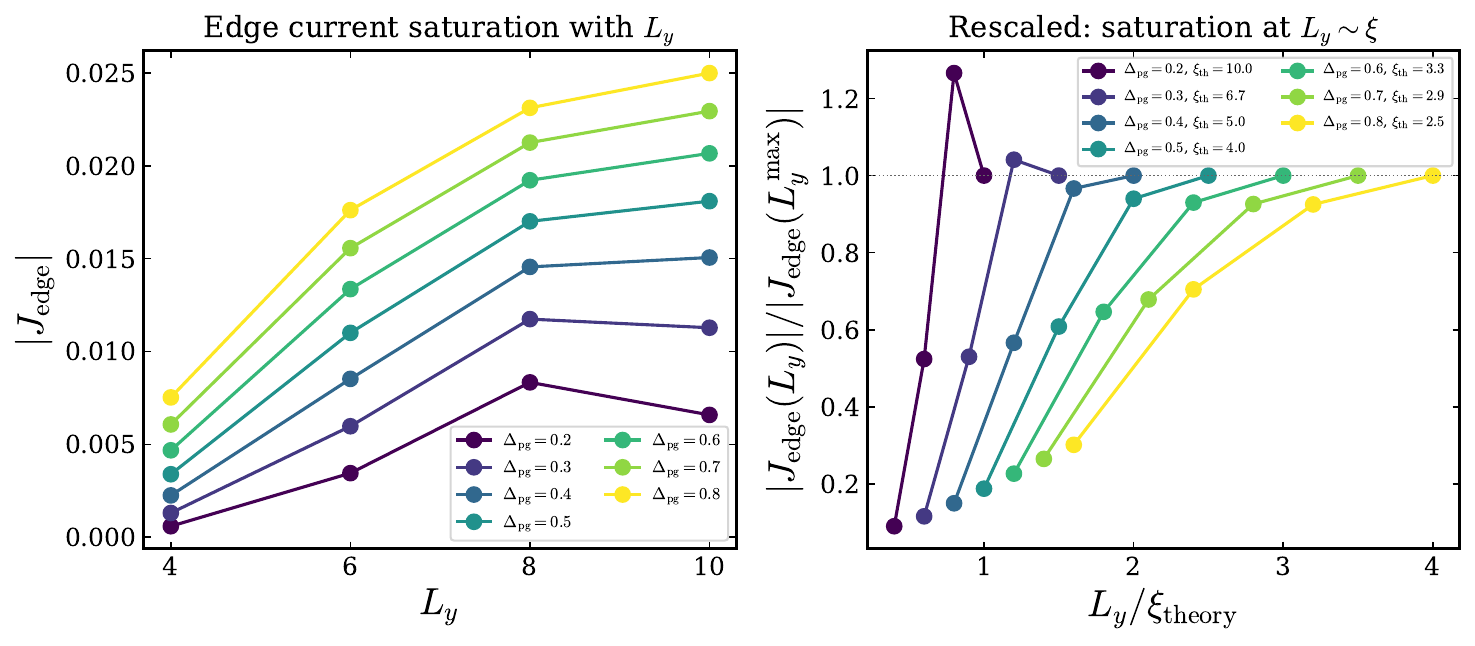}
  \caption{Finite-width edge-current benchmark.
Left: edge-current magnitude against circumference for the imposed
pair amplitudes. Right: the same data against $L_y/\xi_{\rm theory}$,
with $\xi_{\rm theory}=v_F/\Dpg$. The observed approach to saturation
is a finite-cylinder consistency check. The available four widths do
not establish a universal collapse, exclude algebraic corrections,
or prove the continuum $c_1=0$ statement for this observable.}
  \label{fig:edge-saturation}
\end{figure}

\paragraph{Five numerical consistency checks.}
(i) The DMRG energy agrees with exact BdG diagonalization, with
$|\delta E|/N$ from $6.3\times10^{-16}$ at $L_y=4$ to
$3.7\times10^{-6}$ at $L_y=10$.  (ii) The matching BdG matrices have a
particle--hole residual below $3.4\times10^{-14}$.  (iii) The edge currents obey
$J_{\rm left}=-J_{\rm right}$ to numerical precision and have the
expected orientation.  (iv) The nearest-neighbour pair amplitudes
recover the Hamiltonian's relative phase,
Eq.~\eqref{eq:pip-signature}, over all 28 cases.  The MPS ansatz does
not enforce this phase, but the Hamiltonian does, so the result verifies
implementation and convergence.  (v) Edge-current saturation versus
$L_y/\xi_{\rm theory}$ and the area-law mid-cylinder entropy are
consistent with a gapped finite-cylinder state.  These checks are
collected in Figs.~\ref{fig:dmrg-summary} and~\ref{fig:edge-saturation}.

The FHS Chern number evaluated for each parameter set returns
$\CBdG=1$ with integer error below $10^{-14}$.  This is a Bloch-Hamiltonian
classification, not an independent invariant extracted from the MPS.
A genuine many-body demonstration of spontaneous chiral superconductivity
would require a number-conserving interacting Hamiltonian with no explicit
complex pair field, followed by a many-body Chern number, flux pump, or
long-distance pairing-correlation analysis.  We do not claim that result
here.

\paragraph{Chiral edge current.}  For the cylinder geometry we define
\begin{equation}
  J_{\rm left/right}
  = \frac{1}{L_y}\sum_{y}
    2\,\mathrm{Im}\!\left[(-t)\,
    \langle c^\dagger_{x_{\rm edge},y} c_{x_{\rm edge},y+1}\rangle
    \right],
\label{eq:edge-current-def}
\end{equation}
evaluated on the first and last rung.  For this symmetric cylinder, the model predicts opposite edge currents, $J_{\rm left}=-J_{\rm right}$. Chirality alone would not require equal magnitudes for inequivalent boundaries. The relation is not imposed by the MPS ansatz.  It is satisfied to
$|J_L + J_R| \leq 3.6\times10^{-15}$ at $L_y = 4$, with maximum relative imbalance about $1.5\times10^{-6}$ over the complete scan.

\paragraph{Energy benchmark.}  Since the $V_{nn} = 0$ Hamiltonian is
quadratic, the DMRG energy can be compared with exact BdG
diagonalisation on the identical cylinder.  The error per site
degrades from machine precision at $L_y = 4$ to
$3.7\times10^{-6}$ at $L_y = 10$, consistent with increasing numerical difficulty at fixed bond-dimension ceiling $\chi_{\max}=512$. A separate bond-dimension extrapolation would be required for a controlled error estimate at each width.

\paragraph{Correlation length from the energy sequence.}  The continuum
dispersion of Eq.~\eqref{eq:pwave-bdg} near $\bm{k}=0$ is
$\xi_{\bm{k}} \approx tk^2-\mu$, so the Fermi velocity is
$v_F = 2\sqrt{t\mu} = 2$ at $t=\mu=1$ and the predicted correlation
length is $\xi_{\rm theory} = v_F/\Dpg$.  A
ratio-method extraction of $\xi_{\rm fit}$ from consecutive
$L_y = 6, 8, 10$ energy shifts gives
$\xi_{\rm fit}/\xi_{\rm theory} = 0.99$ at $\Dpg = 0.8$, varying across the smaller-gap cases, where the available circumferences may not reach the asymptotic regime (Fig.~\ref{fig:ratio-xi-app}).  This is a three-point
fit and we quote it as a consistency check on the exponential
envelope, not as a precision determination of $\xi$.

\paragraph{Entanglement.}  The mid-cylinder von Neumann entropy
$S_{\rm mid}$ ranges from $0.555$ at $(L_y,\Dpg) = (4,0.2)$ to $1.617$
at $(10,0.8)$.  Its size dependence over the four available
circumferences is consistent with a gapped area law, but the range is
too short to resolve subleading logarithmic or constant terms.  The
chiral central charge is more directly diagnosed by an entanglement
spectrum than by this entropy sequence alone.

Figures~\ref{fig:dmrg-summary} and~\ref{fig:edge-saturation} summarize
these checks and the finite-width edge-current trends.

\section{Supporting numerical data}
\label{app:supporting}

\emph{This appendix collects the detailed per-parameter numerical data
underlying the single-particle validation
(Appendix~\ref{app:single-particle}), the Wilson-loop finite-size scan
(Appendix~\ref{app:wilson-loop}), and the DMRG tensor-network benchmark
(Appendix~\ref{app:dmrg}).  All quantities support the summary
statements in the main text.}

\subsection{FHS Chern number for the square-lattice $p+ip$ anchor}

Table~\ref{tab:pip-chern-app} gives the FHS Chern number of the
square-lattice $p+ip$ model at each pairing amplitude $\Delta_0$ in
the scan of Appendix~\ref{app:single-particle}.  The integer error is
at machine precision throughout, confirming that $\CBdG = 1$ is
independent of the pairing strength in the weak-pairing regime.

\begin{table}[t]
  \centering
  \caption{FHS Chern number for the square-lattice $p+ip$ model
    ($t=1$, $\mu=1$, $N_k = 81$) at each pairing amplitude
    $\Delta_0$.  The integer error is bounded by $10^{-14}$ at every
    point; it is eigensolver roundoff and is not reproducible
    digit-for-digit between runs, so a bound rather than a value is
    quoted.  The last column is the full particle--hole gap
    $2E_{\min}$, a sampled direct band separation, rather than the single-quasiparticle activation gap $E_{\min}$.}
  \label{tab:pip-chern-app}
  \begin{ruledtabular}
  \begin{tabular}{cccc}
    $\Delta_0$ & $C_{\rm BdG}^{\rm FHS}$ & $|C - 1|$ & $2E_{\min}$ \\
    \hline
    0.2 & 1 & $<10^{-14}$ & 0.347 \\
    0.3 & 1 & $<10^{-14}$ & 0.520 \\
    0.4 & 1 & $<10^{-14}$ & 0.694 \\
    0.5 & 1 & $<10^{-14}$ & 0.865 \\
    0.6 & 1 & $<10^{-14}$ & 1.024 \\
    0.7 & 1 & $<10^{-14}$ & 1.185 \\
    0.8 & 1 & $<10^{-14}$ & 1.346 \\
  \end{tabular}
  \end{ruledtabular}
\end{table}

\subsection{Four benchmark cases for the annular Fermi sea}

Table~\ref{tab:four-cases-app} lists the four benchmark cases
underlying the vortex-crossing scan of
Fig.~\ref{fig:even-chern-scan}.  All integer errors are at or below
$8.2\times 10^{-15}$.

\begin{table}[t]
  \centering
  \caption{Direct FHS Chern numbers for the four benchmark cases
    ($r_{\rm in}=0.45$, $r_{\rm out}=1.35$, $q=0.90$, $\Delta_0=0.50$,
    $\chi_\Delta=-1$, $N_k=201$).  $Q_{\rm analytic}$ is the occupied-vortex
    charge from Eq.~\eqref{eq:CBdG-vortex}.  $\max|F_{\rm plaq}|$ is
    the largest plaquette flux encountered; it is smaller than $0.05$ in every case here. Small flux is a useful resolution check, while continuum sector assignments require refinement or an independent analytic criterion.  Residual integer errors are eigensolver
    roundoff and are quoted as a bound.}
  \label{tab:four-cases-app}
  \begin{ruledtabular}
  \begin{tabular}{llccc}
    Fermi sea & Gap texture & $C_{\rm BdG}^{\rm FHS}$ & $Q_{\rm analytic}$ & $\max|F_{\rm plaq}|$ \\
    \hline
    disk     & minimal (central only)       & $+1$ & $+1$ & 0.002 \\
    annular  & minimal (central excluded)   & $\phantom{+}0$ & $\phantom{+}0$ & 0.008 \\
    disk     & central $+$ vortex pair      & $+3$ & $+3$ & 0.032 \\
    annular  & vortex pair inside annulus   & $+2$ & $+2$ & 0.047 \\
  \end{tabular}
  \end{ruledtabular}
  \begin{flushleft}
  \footnotesize
  All integer errors $|C_{\rm BdG}^{\rm FHS} - Q_{\rm analytic}|
  < 10^{-14}$.  Chirality reversal $\chi_\Delta \to +1$ on the fourth row
  gives $C_{\rm BdG}^{\rm FHS} = -2$ at the same precision.
  \end{flushleft}
\end{table}

\subsection{Wilson-loop finite-size scan}

Table~\ref{tab:wilson-app} gives the Wilson-loop flux-threading data
used for the finite-width consistency test in
Appendix~\ref{app:wilson-loop}.  The winding $\Delta P = 1$ is exact
at machine precision for every $L_y$; the raw RMS deviation is
oscillatory. The fitted decay scales can be compared qualitatively with the continuum estimate $\xi=\hbar v_F/\Delta_0=5$.

\begin{table}[H]
  \centering
  \caption{Wilson-loop flux-threading results for the square-lattice
    $p+ip$ model ($t=1$, $\mu=1$, $\Delta_0=0.4$).}
  \label{tab:wilson-app}
  \begin{ruledtabular}
  \begin{tabular}{ccccc}
    $L_y$ & $\Delta P$ & winding error & RMS deviation & max deviation \\
    \hline
    4  & 1.000 & 0.000 & 0.0448 & 0.0682 \\
    6  & 1.000 & 0.000 & 0.0277 & 0.0398 \\
    8  & 1.000 & 0.000 & 0.0064 & 0.0103 \\
    10 & 1.000 & 0.000 & 0.0149 & 0.0214 \\
    12 & 1.000 & 0.000 & 0.0068 & 0.0097 \\
    14 & 1.000 & 0.000 & 0.0023 & 0.0033 \\
  \end{tabular}
  \end{ruledtabular}
\end{table}

\subsection{DMRG per-$L_y$ summary}

Table~\ref{tab:dmrg-summary-app} tabulates the DMRG summary
statistics: extremal errors across the seven $\Dpg$ at each $L_y$.

\begin{table}[H]
  \centering
  \caption{DMRG summary for $L_x = 8$, $L_y \in \{4, 6, 8, 10\}$,
    $\Dpg \in \{0.2, \ldots, 0.8\}$, $t = \mu = 1$,
    $\chi_{\max} = 512$, $V_{nn} = 0$.  All errors are maxima over
    the seven $\Dpg$ at each $L_y$.}
  \label{tab:dmrg-summary-app}
  \begin{ruledtabular}
  \begin{tabular}{ccccccc}
    $L_y$ & $|C-1|$ & $|\delta E|/N$ & $|J_L+J_R|$ &
    PH-sym err & $|\phi_y-\phi_x+\pi/2|$ & $S_{\rm mid}$ range \\
    \hline
     4 & $2.8\times 10^{-15}$ & $6.3\times 10^{-16}$ &
         $3.6\times 10^{-15}$ & $2.2\times 10^{-14}$ &
         $2.2\times 10^{-14}$ & 0.555--0.800 \\
     6 & $2.8\times 10^{-15}$ & $1.9\times 10^{-11}$ &
         $5.2\times 10^{-12}$ & $2.8\times 10^{-14}$ &
         $1.8\times 10^{-11}$ & 0.820--1.074 \\
     8 & $2.0\times 10^{-15}$ & $5.5\times 10^{-8\phantom{0}}$ &
         $2.6\times 10^{-9\phantom{0}}$ & $3.2\times 10^{-14}$ &
         $4.4\times 10^{-9\phantom{0}}$ & 1.365--1.550 \\
    10 & $2.9\times 10^{-15}$ & $3.7\times 10^{-6\phantom{0}}$ &
         $2.7\times 10^{-8\phantom{0}}$ & $3.4\times 10^{-14}$ &
         $2.8\times 10^{-7\phantom{0}}$ & 1.548--1.617 \\
  \end{tabular}
  \end{ruledtabular}
\end{table}

\subsection{Ratio-method correlation length from DMRG energies}

\begin{figure}[t]
  \centering
  \includegraphics[width=0.55\textwidth]{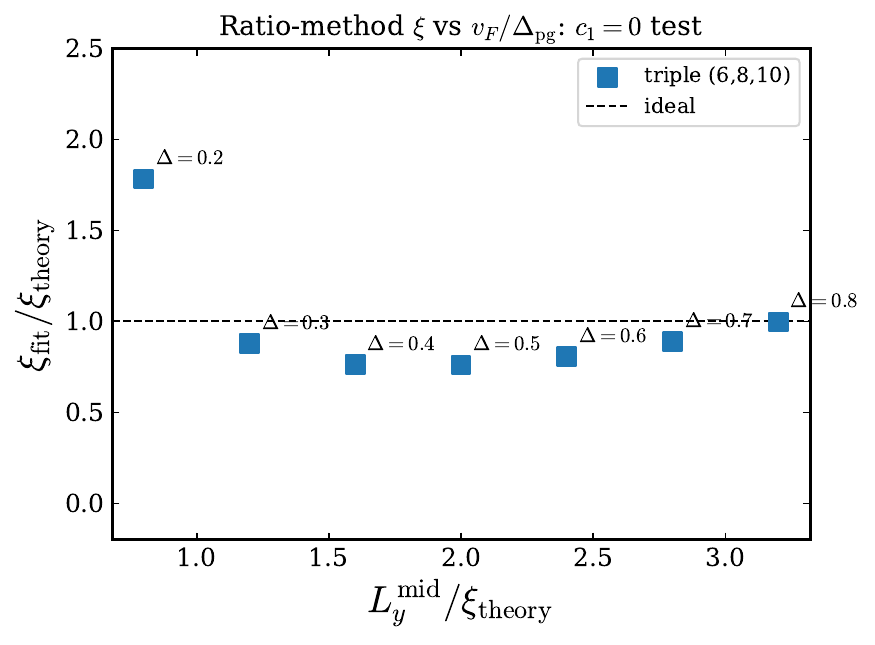}
  \caption{Three-width energy-ratio estimate of a decay length.
The consecutive widths are $6,8,10$. For a single exponential,
$\xi_{\rm fit}=-2/\ln[(e_{10}-e_8)/(e_8-e_6)]$.
The ratio to $v_F/\Dpg$ is about $0.99$ at $\Dpg=0.8$.
Intermediate-gap ratios are below unity and the smallest-gap ratio is
above it. Three points supply an effective scale but do not select an
asymptotic functional form or establish a precision correlation length.}
  \label{fig:ratio-xi-app}
\end{figure}

Figure~\ref{fig:ratio-xi-app} extracts the correlation length
$\xi_{\rm fit}$ from consecutive-$L_y$ energy shifts and compares it
to the continuum prediction $\xi_{\rm theory} = v_F/\Dpg$.  At
$\Dpg = 0.8$ (rightmost point) the ratio
$\xi_{\rm fit}/\xi_{\rm theory} = 0.99$; the intermediate-gap ratios lie below unity, while the smallest-gap value rises above it. The accessible widths need not lie in the asymptotic regime.  The convergence toward unity for
$L_y^{\rm mid}/\xi_{\rm theory} \gtrsim 1$ is consistent with the same
exponential finite-cylinder scale tested by the Wilson-loop calculation
of Appendix~\ref{app:wilson-loop}.

\end{document}